\documentclass[prb,aps, twocolumn, amsmath,amssymb,superscriptaddress]{revtex4}
\usepackage{float}
\usepackage{amsmath}
\usepackage{amssymb}
\usepackage{amsfonts}
\usepackage{euscript}
\usepackage{enumerate}
\usepackage{hhline}
\usepackage{pslatex}
\usepackage{tabularx}
\usepackage{threeparttable}
\usepackage[usenames,dvipsnames]{xcolor}
\usepackage[colorlinks,linkcolor=red,anchorcolor=blue,citecolor=blue]{hyperref}

\usepackage{graphicx}
\usepackage{dcolumn}
\usepackage{bm}
\usepackage[sort&compress]{natbib}

\makeatletter

\newcommand{\bra}[1]{\langle #1 | }
\newcommand{\ket}[1]{|  #1 \rangle }

\renewcommand{\@biblabel}[1]{#1. }
\renewcommand{\@dotsep}{500}
\renewcommand{\@pnumwidth}{0em}
\renewcommand{\l@figure}[2]{
\@dottedtocline{1}{1.5em}{2em}{Figure #1}{}\vspace{15pt}}

\begin{document}

\title{Tailoring solid-state single-photon sources with stimulated emissions}

\author{Yuming Wei}
\thanks{These authors contributed equally}
\affiliation{State Key Laboratory of Optoelectronic Materials and Technologies, School of Physics, School of Electronics and Information Technology, Sun Yat-sen University, Guangzhou 510275, China}

\author{Shunfa Liu}
\thanks{These authors contributed equally}
\affiliation{State Key Laboratory of Optoelectronic Materials and Technologies, School of Physics, School of Electronics and Information Technology, Sun Yat-sen University, Guangzhou 510275, China}

\author{Xueshi Li}
\affiliation{State Key Laboratory of Optoelectronic Materials and Technologies, School of Physics, School of Electronics and Information Technology, Sun Yat-sen University, Guangzhou 510275, China}

\author{Ying Yu}
\affiliation{State Key Laboratory of Optoelectronic Materials and Technologies, School of Physics, School of Electronics and Information Technology, Sun Yat-sen University, Guangzhou 510275, China}

\author{Xiangbin Su}
\affiliation{State Key Laboratory for Superlattice and Microstructures, Institute of Semiconductors, Chinese Academy of Sciences, Beijing 100083, China.}
\affiliation{Center of Materials Science and Optoelectronics Engineering, University of Chinese Academy of Sciences, Beijing 100049, China.}

\author{Shulun Li}
\affiliation{State Key Laboratory for Superlattice and Microstructures, Institute of Semiconductors, Chinese Academy of Sciences, Beijing 100083, China.}
\affiliation{Center of Materials Science and Optoelectronics Engineering, University of Chinese Academy of Sciences, Beijing 100049, China.}

\author{Xiangjun Shang}
\affiliation{State Key Laboratory for Superlattice and Microstructures, Institute of Semiconductors, Chinese Academy of Sciences, Beijing 100083, China.}
\affiliation{Center of Materials Science and Optoelectronics Engineering, University of Chinese Academy of Sciences, Beijing 100049, China.}

\author{Hanqing Liu}
\affiliation{State Key Laboratory for Superlattice and Microstructures, Institute of Semiconductors, Chinese Academy of Sciences, Beijing 100083, China.}
\affiliation{Center of Materials Science and Optoelectronics Engineering, University of Chinese Academy of Sciences, Beijing 100049, China.}

\author{Huiming Hao}
\affiliation{State Key Laboratory for Superlattice and Microstructures, Institute of Semiconductors, Chinese Academy of Sciences, Beijing 100083, China.}
\affiliation{Center of Materials Science and Optoelectronics Engineering, University of Chinese Academy of Sciences, Beijing 100049, China.}

\author{Haiqiao Ni}
\affiliation{State Key Laboratory for Superlattice and Microstructures, Institute of Semiconductors, Chinese Academy of Sciences, Beijing 100083, China.}
\affiliation{Center of Materials Science and Optoelectronics Engineering, University of Chinese Academy of Sciences, Beijing 100049, China.}

\author{Siyuan Yu}
\affiliation{State Key Laboratory of Optoelectronic Materials and Technologies, School of Physics, School of Electronics and Information Technology, Sun Yat-sen University, Guangzhou 510275, China}

\author{Zhichuan Niu}
\affiliation{State Key Laboratory for Superlattice and Microstructures, Institute of Semiconductors, Chinese Academy of Sciences, Beijing 100083, China.}
\affiliation{Center of Materials Science and Optoelectronics Engineering, University of Chinese Academy of Sciences, Beijing 100049, China.}

\author{Jake Iles-Smith}
\affiliation{Department of Physics and Astronomy, The University of Manchester, Oxford Road, Manchester, M13 9PL, UK
}
\affiliation{Department of Electrical and Electronic Engineering, The University of Manchester, Sackville Street Building, Manchester, M1 3BB, UK
}

\author{Jin Liu}
\thanks{liujin23@mail.sysu.edu.cn}
\affiliation{State Key Laboratory of Optoelectronic Materials and Technologies, School of Physics, School of Electronics and Information Technology, Sun Yat-sen University, Guangzhou 510275, China}

\author{Xuehua Wang}
\affiliation{State Key Laboratory of Optoelectronic Materials and Technologies, School of Physics, School of Electronics and Information Technology, Sun Yat-sen University, Guangzhou 510275, China}

\date{\today}

\begin{abstract}
\noindent \textbf{The coherent interaction of electromagnetic fields with solid-state two-level systems can yield deterministic quantum light sources for photonic quantum technologies. To date, the performance of semiconductor single-photon sources based on three-level systems is limited mainly due to a lack of high photon indistinguishability. Here, we tailor the cavity-enhanced spontaneous emission from a ladder-type three-level system in a single epitaxial quantum dot (QD) through stimulated emission. After populating the biexciton (XX) of the QD through two-photon resonant excitation (TPE), we use another laser pulse to selectively depopulate the XX state into an exciton (X) state with a predefined polarization. The stimulated XX-X emission modifies the X decay dynamics and yields improved polarized single-photon source characteristics such as a source brightness of 0.030(2), a single-photon purity of 0.998(1), and an indistinguishability of 0.926(4). Our method can be readily applied to existing QD single-photon sources and expands the capabilities of three-level systems for advanced quantum photonic functionalities.}
\end{abstract}

\maketitle

Interactions between quantum emitters and electromagnetic fields 
are fundamental to
atomic physics\cite{drake2006atomic} and quantum optics\cite{fox2006quantum}, with widespread applications in 
modern quantum technologies. 
Coherent excitations of a two-level-system were initially realized in atoms with 
milestone observations of photon anti-bunching in resonance fluorescence, the Mollow triplet, and Rabi oscillation\cite{fox2006quantum}. 
In solid-state systems, e.g., semiconductor QDs, the coherent manipulation of two-level systems not only boosts the developments of high-performance chip-scale quantum light sources\cite{senellart2017high} and deterministic photon-emitter interfaces\cite{uppu2021quantum}, but also enables the experimental demonstration of long sought-after theoretical predictions in atomic physics\cite{schulte2015quadrature,he2015dynamically}. Extending a two-level system to a three-level system brings enormous opportunities for both exploring fundamental physics and device applications. For example, the spontaneous Raman scattering process in $\Lambda$-type three-level systems is employed to tune spectral compositions and temporal wave packets of the single photons emitted from cold atoms\cite{chou2004single}, trap ions\cite{almendros2009bandwidth}, and semiconductor QDs\cite{pursley2018picosecond}, while the stimulated Raman adiabatic passage is exploited to deterministically prepare specific quantum states of both quantum emitters\cite{pillet1993adiabatic} and photons\cite{kuhn2002deterministic}. However, the realization of a $\Lambda$-type three-level system in QDs requires a high magnetic field (typically a few Tesla) to introduce pronounced Zeeman splittings of charged exciton states\cite{xu2008coherent}. Alternatively, ladder-type three-level systems can be prepared optically by addressing the cascaded XX-X decay process in single QDs\cite{moreau2001quantum}. Due to the carrier confinements in QDs, the enhanced Coulomb interactions shift the energy of XX from that of X by a binding energy of a few~meV\cite{hu1990biexcitons}. The XX and X form an anharmonic ladder-type three-level system, facilitating the realizations of a serial of solid-state quantum photonic devices, e.g., quantum logic gates\cite{li2003all} and single-photon transistors\cite{hwang2009single}. Furthermore, unexplored physics in QDs could be revealed by depopulating excited states of QDs in both coherent\cite{spinnler2021optically} and incoherent\cite{liu2016ultrafast} ways, which may results in novel applications such as single-particle tracking\cite{piatkowski2019ultrafast} and super-resolution\cite{kianinia2018all,kaldewey2018far}.

In this work, we employed two-photon resonant excitation (TPE) and stimulated emission to coherently populate and de-populate a ladder-type three-level system in order to improve the polarized single-photon emission from a single QD 
deterministically coupled to a micropillar cavity~\cite{Liu2021}.
TPE is used to directly populate the XX from the ground state\cite{muller2014demand}, and a subsequent stimulating pulse selectively and coherently de-populates the XX state to a desired X state coupled to the cavity mode of the micropillar. 
This is in contrast to Raman processes, where two detuned lasers are employed to coherently transfer population, but the excited state is never populated. By avoiding the time jitter associated with the incoherent spontaneous emission of XX and suppressing the generation of single-photons with the unwanted polarization, the device performances as a polarized single-photon source are significantly boosted compared to the conventional TPE. Specifically, the in-fiber efficiency, single photon purity and indistinguishability are improved from 0.016(2), 0.969(4), 0.27(3) to 0.030(2), 0.998(1) and 0.926(4) respectively. The in-fiber efficiency of this source could be improved to 0.51(1) with optimized optics and non-blinking QDs. Despite the improved brightness of polarized single-photon emission, the limit of the achievable photon indistinguishablity with our method is still set by strict resonant excitation. There are alternative techniques suppressing the photon loss in resonant excitation\cite{wang2019towards,huber2020filter}, however the employment of stimulated emission doesn’t rely on specific geometries of photonic nanostructures and therefore it could serve as an alternative for QD-based devices to reach high-performance single-photon sources without involving resonance fluorescence. With further improvements in the QD quality and optimizations of the photonic nanostructures, optimal single-photon sources may be anticipated with the assistance of stimulated emissions.

\begin{center}
	\begin{figure}
		\begin{center}
			\includegraphics[width=0.98\linewidth]{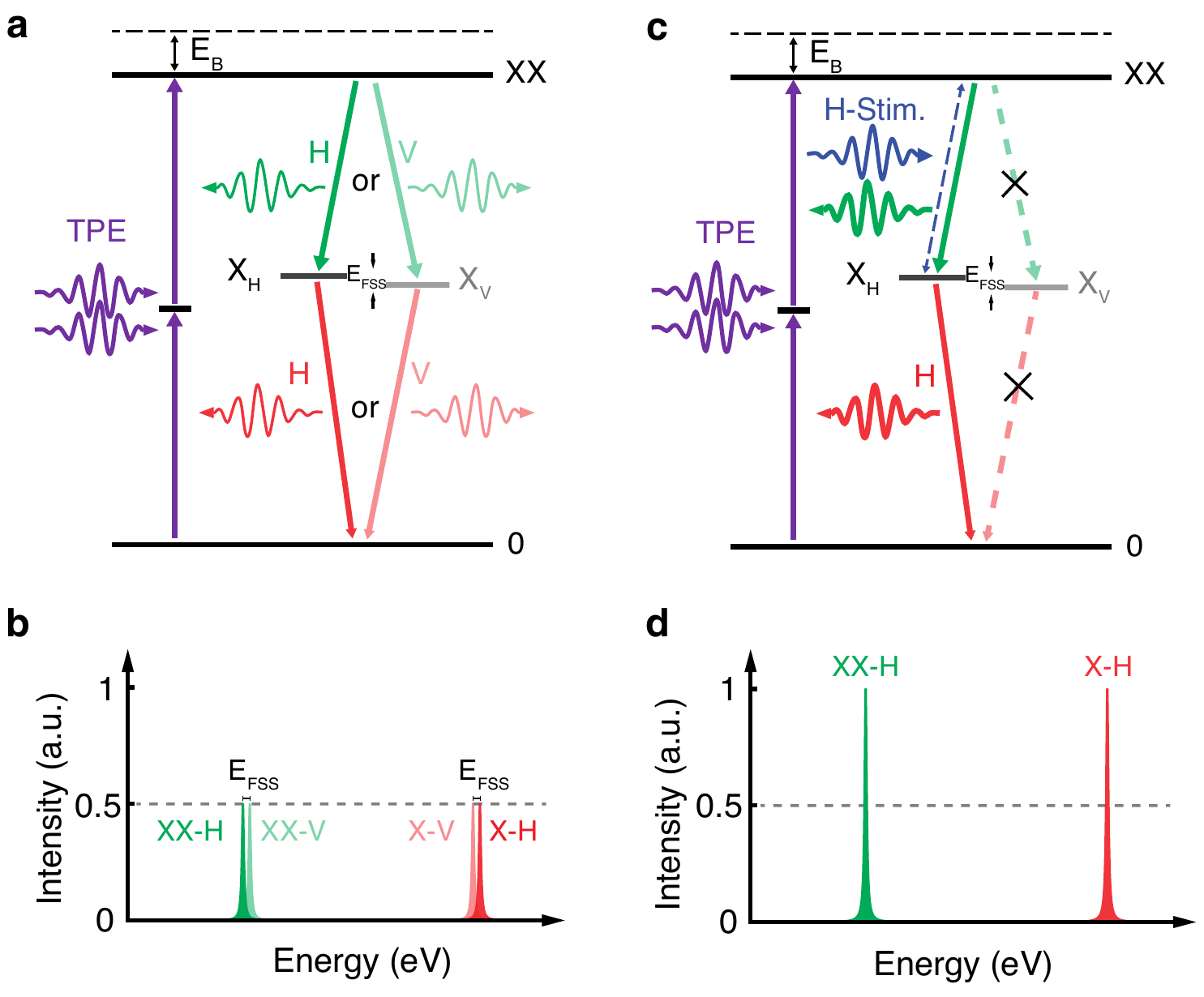}
			\caption{\textbf{Operation principle of TPE and TPE with stimulating lasers.} (a) Schematic of the TPE process. XX states are firstly populated by using TPE pulses (purple arrows). The XX states randomly decay to the $\rm X_H$ or $\rm X_V$ by emitting H-polarized (green arrow) or V-polarized (light green arrow) single photons respectively. The $\rm X_H$ and $\rm X_V$  then relax to the ground state by emitting the second H-polarized (red arrow) or V-polarized (light red arrow) single photons. (b) Schematic of the spectrum associated to the TPE process. Both XX and X exhibit H-polarized and V-polarized emission lines separated by $\rm E_{FSS}$. (c) Schematics of TPE with stimulating lasers. Once the XX states are populated by TPE pluses, stimulating pulses (blue arrow) selectively bring the XX to $\rm X_H$ via stimulated emissions of H-polarized photons, which greatly suppresses the formation of $\rm X_V$. The $\rm X_H$ states then decays to the ground state by emitting the second H-polarized photons. (d) Schematic of the spectrum associated to the TPE with stimulating lasers. The X and XX only show H-polarized emission lines with two fold intensities compared with the TPE case in (b). The V-polarized single-photons can also be generated if switching the polarization of the stimulating lasers to V, as demonstrated in Fig.~3(b).}
			\label{fig:Fig1}
		\end{center}
	\end{figure}
\end{center}

\section{Operation principles}

The details of device fabrication and simulation are presented in Supplementary Information Section $\rm \uppercase\expandafter{\romannumeral1}$ and Section $\rm \uppercase\expandafter{\romannumeral2}$. Fig.~1 shows the operation principles of both TPE and TPE with stimulating lasers. In Fig.~1(a), TPE pulses are firstly employed to prepare XX states. Without the introduction of the simulating pulses, the XX state randomly decays to H-polarized exciton ($\rm X_H$) or V-polarzied exciton ($\rm X_V$) by emitting either H-polarized or V-polarized single photons. 
The X states then decay to the ground state by emitting 
a second single photon,with the same polarization as the first due to selection rules.
One of the advantages of the TPE is to suppress the re-excitation process and therefore enables the realization of an ultra-high single-photon purity~\cite{schweickert2018demand} of $g^{(2)}(0) = (7.5\pm 1.6)\times 10^{-5} $. However, due to the time jitter associated to the spontaneous emission of XX, the indistinguishability of photons emitted by X is inherently limited by $P = \gamma_{XX}/(\gamma_{XX}+\gamma_{X})$, where $\gamma_{XX}$ and $\gamma_{X}$ are the decay rates of XX and X respectively~\cite{scholl2020crux}. In addition, the polarization state of the emitted single photons from both X and XX are a mixture of H- and V-polarization as schematically shown in Fig.~1(b), reducing the source brightness by a factor of two when working as a polarized single-photon source.
Although our focus in this work is the generation of polarized single-photon emission for specific quantum simulation proposals, e.g. Boson sampling~\cite{brod2019photonic}, un-polarized single-photons sources are also important in the context of photonic quantum technologies with applications in quantum metrology\cite{giovannetti2006quantum}, random number generation\cite{luo2020quantum}, and quantum lithography\cite{boto2000quantum}.

To overcome the fundamental limit of the indistinguishability and the random polarization of single photons emitted by X, we introduce H-polarized pulses with the same energy as the XX right after the TPE pulses, as presented in Fig.~1(c). The H-polarized stimulating pulses instantaneously transfers population from the XX to $\rm X_H$, preventing the formation of $\rm X_V$, which doubles the intensity of the H-polarized single photons emitted by X, as shown in Fig.~1(d). Similarly, the V-polarized single photons are generated with V-polarized stimulating lasers, which is shown in Fig.~3(b). 
The optical properties of the QD can be understood through a simple four-level model, where we find an analytic expression for the indistinguishability of the form
\begin{equation}\label{eq:ind}
I = I_\mathrm{res} \frac{2\gamma_{XX} + \gamma_\mathrm{stim}}{2\gamma_{XX} +  F_\mathrm{P}\gamma_{X}+\gamma_\mathrm{stim}},
\end{equation}
where $\gamma_\mathrm{stim}$ is the stimulated emission rate, and $F_\mathrm{P}$ is the Purcell factor of the $\rm X_V \rightarrow$0 transition. 
The above expression can be factorised into two contributions: the indistinguishability that would be observed under resonant excitation~\cite{kaer2013role} given by $I_\mathrm{res} = F_\mathrm{P}\gamma_{X}/( F_\mathrm{P}\gamma_{X} + \Gamma)$, where $\Gamma$ is the pure dephasing rate of the QD, and a second factor that reduces the observed indistinguishability due to the time jitter. 
In the absence of the stimulated process ($\gamma_\mathrm{stim}=0$), the indistinguishability is reduced to its minimum value, as a consequence of time-jitter~\cite{unsleber2015two}.
Increasing $\gamma_\mathrm{stim}$  monotonically increases the indistinguishability, until the limit $\gamma_\mathrm{stim}$ dominates over the spontaneous emission rates, where it reaches idealised limit of strict resonant excitation $I\approx I_\mathrm{res}$. 
A detailed derivation and analysis of this model is provided in the Supplementary Information Section $\rm \uppercase\expandafter{\romannumeral3}$.

\begin{figure*}
	\includegraphics[width=0.95\linewidth]{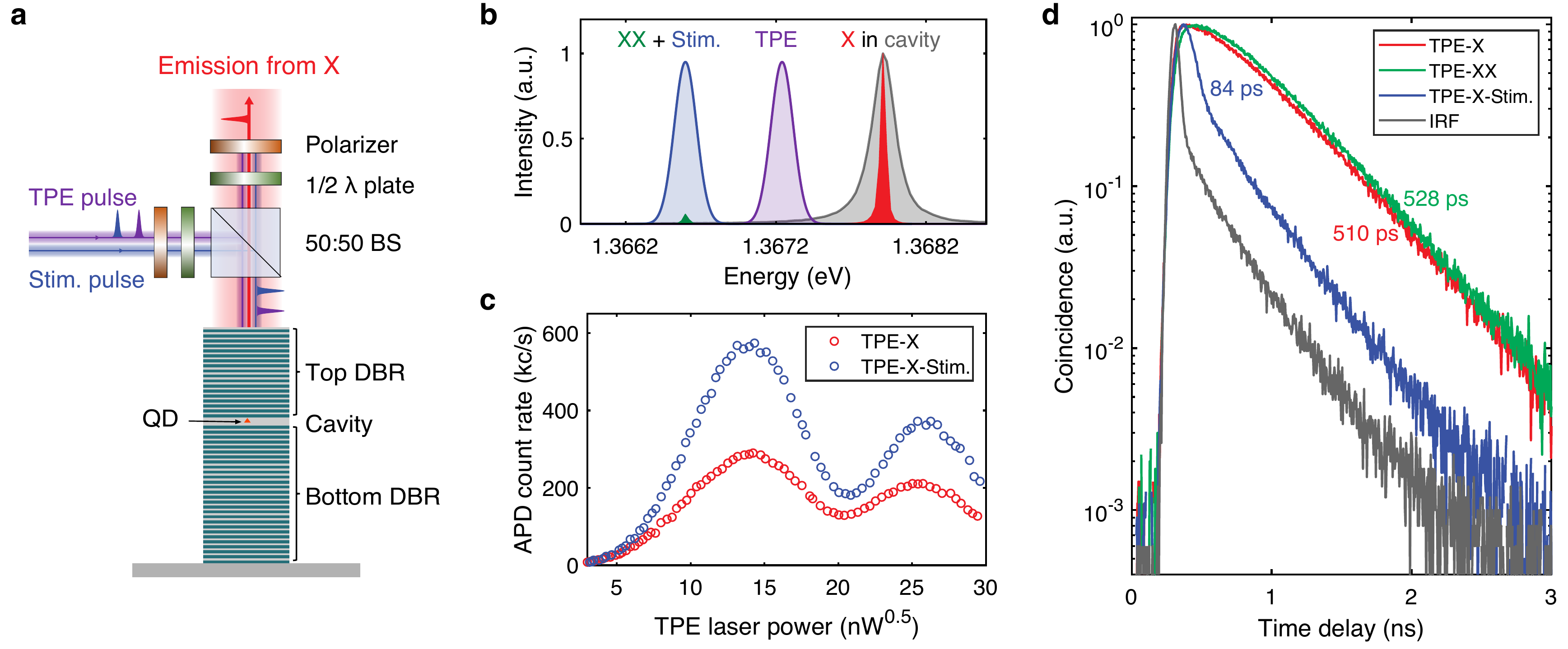}
	\caption{\textbf{Purcell enhanced spontaneous emission mediated by stimulated emission.} (a) Schematic of the experiments. TPE and stimulating pulse sequences are exploited to excite the single QD in a micropillar and the targeted emissions are collected from the top of the device. The half-wave plates and polarizers are used to control the polarizations of excitation laser and the emitted single photons. BS: beamsplitter, DBR: distributed Bragg reflector. (b) Spectrum of coupled QD-micropillar system under TPE. The X (red) is blue-detuned from the XX (green) and resonant with the cavity mode of the micropillar (light black). The TPE (light purple) pulse are spectrally located in the middle of X and XX. The stimulating pulses (light blue) have the same energy as the XX. (c) The intensity of H-polarized single photons from X as a function of the TPE laser power with (blue) and without (red) H-polarized stimulating pulses. X exhibits characteristic Rabi oscillations and the stimulating pluses double the intensity of the H-polarized single photons. (d). Lifetimes of X and XX with and without the stimulating pulses. X and XX show very similar lifetimes of 510 and 528 ps respectively under TPE. The lifetime of X is shortened to 84 ps with the introduction of the stimulating pulses.}
	\label{fig:Fig2}
\end{figure*}

\begin{figure*}
	\begin{center}
		\includegraphics[width=0.95\linewidth]{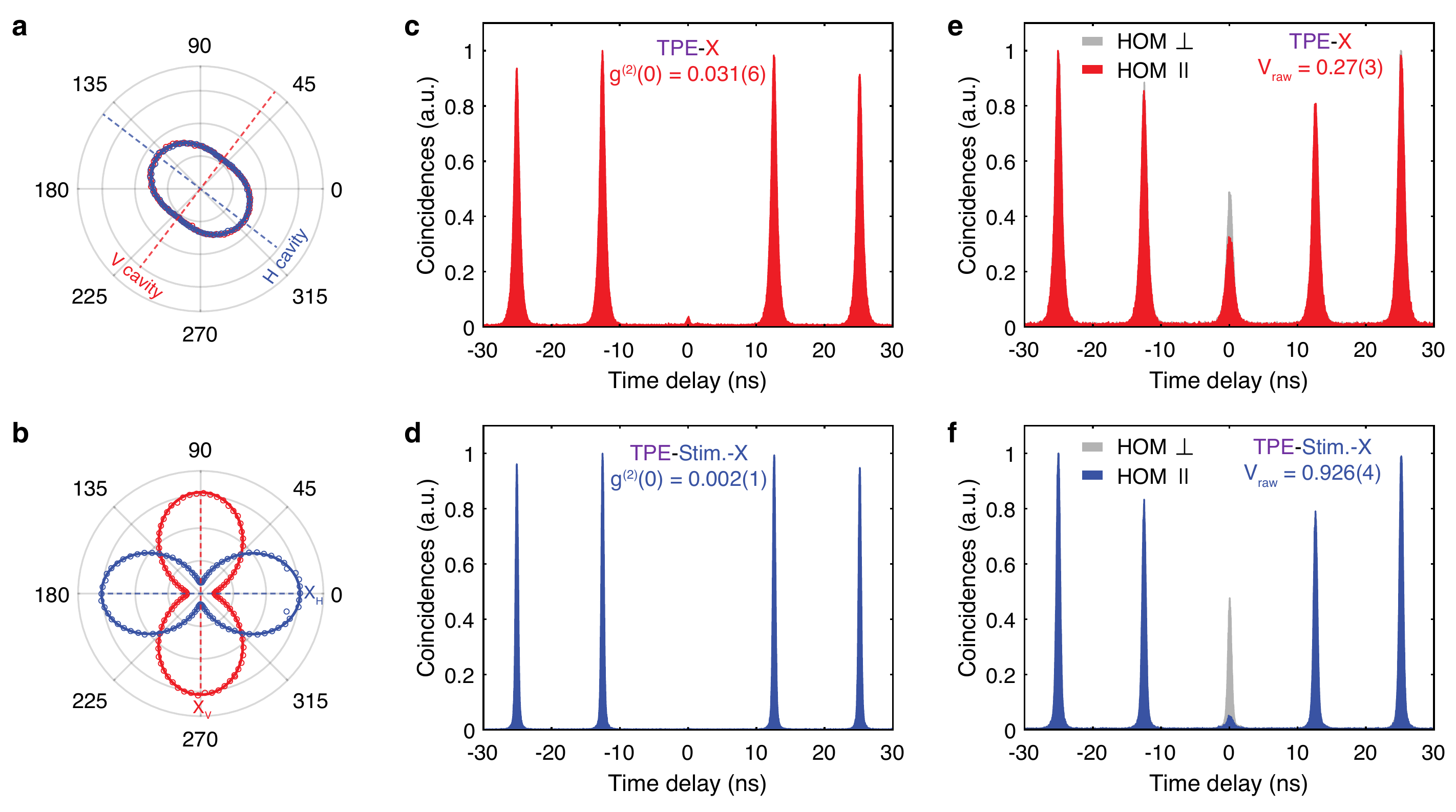}
		\caption{\textbf{Improved performances of the QD-micropillar single-photon source with stimulated emissions.} (a) Polarization states of the single photons excited by TPE pulses with H- and V-polarization. The emitted photons are weakly H-polarized with a degree of polarization (DOP) as low as 0.18 because X couples to the H-polarized cavity mode slightly better than the V-Polarized cavity mode (see Supplementary Information Section $\rm \uppercase\expandafter{\romannumeral5}$.). The polarization axis of the cavity modes are denoted by the dash lines. The polarizations of the emitted single photons are not dependent on the polarization of the TPE pulses. (b) Polarization states of the single photons excited by TPE pulses followed by H-and V-polarized stimulating pulses. The single photons are highly-polarized and share the same polarization states as the stimulating pulses. The DOPs of H-polarized and V-polarized photons are 0.81 and 0.75 respectively. In addition, the intensities of polarized single photons are almost as twice as those without stimulating pulses as shown in (a). (c) Second-order correlation of the single photons from X under TPE. (d) Second-order correlation of the single photons from X under TPE followed by stimulating pulses. The $g^{(2)}(0)$ value is lowered from 0.031(6) to 0.002(1) with the assistance of stimulating pulses. The uncertainty is a one standard deviation value of a double-Lorentzian fit to the peak area of the histogram at zero delay. (e) Hong-Ou-Mandel (HOM) measurement of the single photons from X under TPE. (f) HOM measurement of the single photons from X under TPE followed by stimulating pulses. On both panels, the colored plots correspond to the HOM interference histograms of the parallel-polarized single photons and the grey graphs correspond to the HOM interference of cross-polarized single photons. The grey area at zero delays are used to quantify the central peak suppression and to extract the visibilities of the HOM interference. The indistinguishability is improved from 0.27(3) to 0.926(4) by removing the time jitter associated to the spontaneous emission of XX with stimulating pulses. The delay time in the HOM interferometer is chosen to be 12.56 ns to match the repetition rate of the excitation laser. The visibilities of the two-photon interference are extracted from the raw data without any corrections, and the uncertainty is a one standard deviation value of a double-Lorentzian fit to the peak area of the histogram at zero delay. All the data are recorded under the $\pi$-pulse power.}
		\label{fig:Fig3}
	\end{center}
\end{figure*}

The TPE with stimulating pulses are applied to a QD-micropillar system and the emission from X is collected from the top of the micropillar, as schematically shown in Fig.~2(a). The full experimental setup is shown in Extended Data Fig.~1. Fig.~2(b) presents the measured photoluminescence spectrum in which the X and XX are at 1.3679~eV and 1.3666~eV respectively under the TPE. The X is resonant with the cavity mode of the micropillar and the stimulating laser is tuned to the XX-X$_\mathrm{H}$ transition. The emission intensity of X is several orders of magnitude higher than XX due to the fact that only emission from the X is enhanced by the cavity mode. The emission intensity of $\rm X_H$ as a function of the TPE laser power with and without the H-polarized stimulating pulses are presented in Fig.~2(c). Under TPE, X can be coherently addressed as a two-level system, exhibiting a characteristic Rabi oscillation~\cite{muller2014demand}. By introducing the stimulating laser pulses with H-polarization, the intensity of the H-polarized single photon emission is doubled while maintaining the feature of Rabi oscillation. The doubling of the H-polarized single photons is attributed to the suppression of the decay path from XX to $\rm X_V$ due to the presence of stimulating pulses in H-polarized state. We measure the decay curves of X under TPE with and without the simulating laser pulses, as presented in Fig.~2(d). Instead of a pronounced enhancement of the decay rate of X expected from the Purcell-effect, X exhibits a relatively long lifetime of 510 ps under TPE. This is because X can only emit single photons after the completion of XX decay process that is not enhanced by the cavity mode (The measured lifetime of XX is 528~ps). With the assistance of the stimulating pulses, the decay bottleneck of XX is removed and a much faster lifetime of 84 ps for X is obtained, revealing an appreciable Purcell effect. A Purcell factor of 6.6 is extracted by comparing the lifetimes of the X on resonance and far off resonance under the resonant excitation, see more details in Supplementary Information Section $\rm \uppercase\expandafter{\romannumeral4}$.

\begin{center}
	\begin{figure}
		\includegraphics[width=\linewidth]{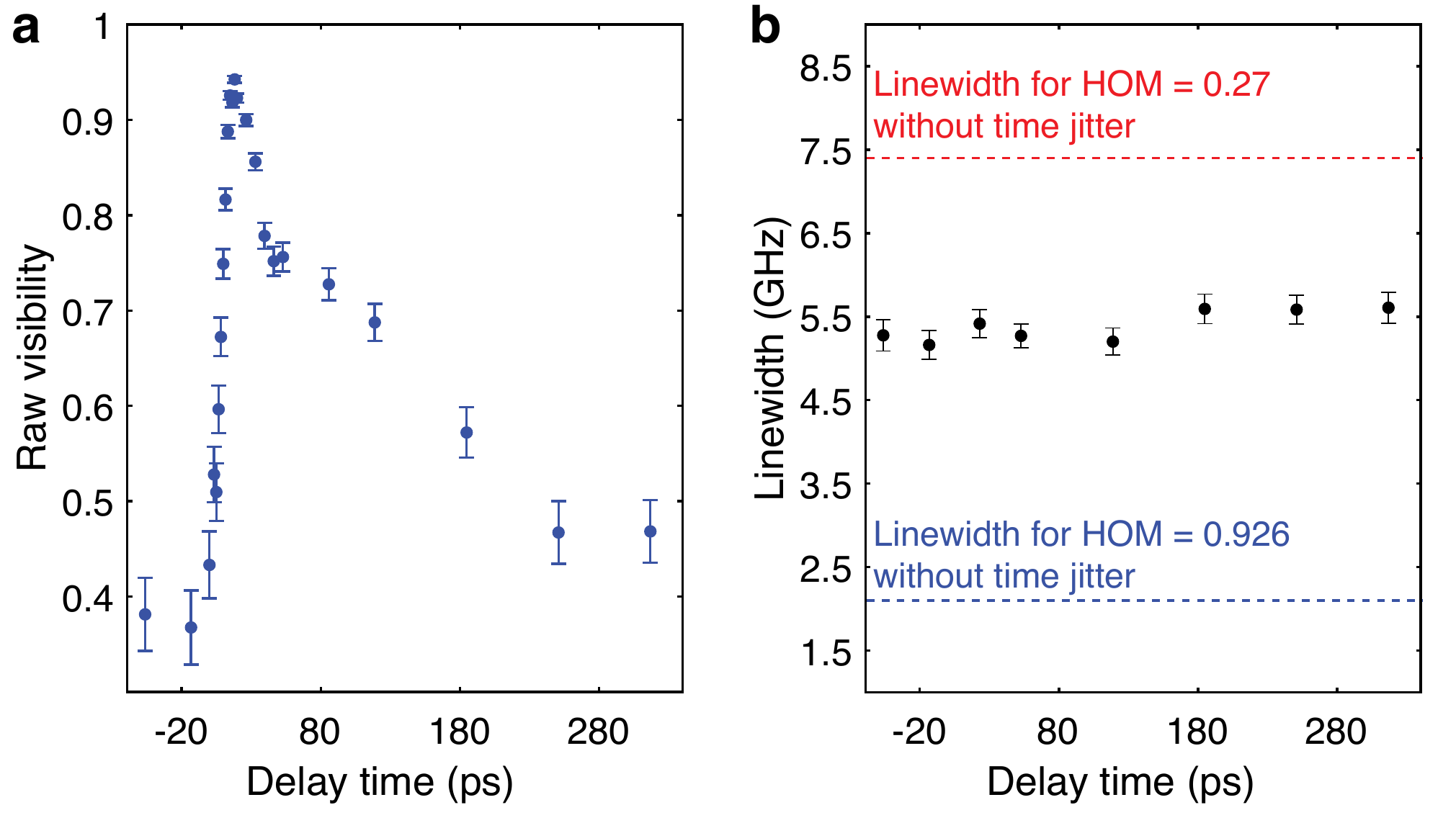}
		\caption{\textbf{Indistinguishability and linewidth as a function of the delay time between the stimulating lasers and the TPE pulses}. (a) Indistinguishability as a function of the delay time between the stimulating lasers and the TPE pulses. The visibilities of the two-photon interference are extracted from the raw data without any corrections, and the uncertainty is a one standard deviation value of a double-Lorentzian fit to the peak area of the HOM histogram at zero delay. (b) Linewidth as a function of the delay time between the stimulating lasers and the TPE pulses. The linewidth is extracted from a Lorentzian fit to the data measured by a scanning Farby-Perot interferometer, and the uncertainty is a one standard deviation value of the fit. Blue line: expected linewidth of a 2.1~GHz X with an indistinguishability of 0.927 (with no emission time jitter). Red line: expected linewidth of a 7.4~GHz X with an indistinguishability of 0.27 (with no emission time jitter).}
		\label{fig:Fig4}
	\end{figure}
\end{center}

\section{Improved device performance as a polarized single-photon source}

We examine the performance of our device as a polarized  single-photon sources with the assistance of the stimulating pulses. Under TPE, a degree of the polarization (DOP) as low as 0.18 is obtained because X couples to the H-polarized cavity mode slightly stronger than the V-polarized cavity mode (see Supplementary Information Section $\rm \uppercase\expandafter{\romannumeral5}$). The polarizations of the photons emitted by X is not dependent on the polarization of the TPE laser, as shown in Fig.~3(a). On the contrary, the stimulating pulses enable the realization of highly-polarized single photons and double the emission intensities, as presented in Fig.~3(b). In addition, the polarization states of the single photons strictly follow the polarizations of stimulating pulses, providing an extra degree of freedom to engineer the polarization states of single-photon sources with ultra-fast laser pulses. The single-photon purity of the emitted photons with and without the stimulating pulses are compared by measuring the second-order correlation as shown in Fig.~3(c,d). The coincidence events at zero delay is completely suppressed with the stimulating pluses, lowering  down the $g^{(2)}(0)$ from 0.031(6) to as low as 0.002(1). 
We attribute the improved single-photon purity in the presence of stimulated emission to an increase in the ratio of population created by direct XX-X transitions (see Supplementary Fig.~S6), versus indirect excitation of the X state through LA-phonon mediated processes~\cite{thomas2021bright}. 
Photons emitted from such phonon-mediated excitation mechanisms have been observed to contain significant multi-photon components ($g^{(2)}(0)=0.03$ as shown in the Table 1).
Due to the avoidance of the emission time jitter associated with spontaneous emission of XX, the raw indistinguishability of the X photons is significantly improved from 0.27(3) to 0.926(4) 
which is still slightly lower than the value obtained under resonant excitation (0.940(3)), as compared in Fig.~3(e,f). The unusual-low indistinguishability of 0.27(3) obtained without the stimulating pulses is due to the pronounced Purcell-factor which shortens the X lifetime and therefore effectively amplifies the time jitter associated with the lifetime of XX, which is in a very good agreement with the expression in Eq.~\ref{eq:ind}. To identify the contribution of the time jitter to the photon indistinguishability, we plot both indistinguishability and linewidth as a function of the delay time between the H-polarized stimulating pulses and TPE pulses. The indistinguishability rises up sharply when the stimulating pulses overlap with TPE pulses and starts to drop as the delay time increases, as shown in Fig.~4(a). The delay time controls the emission time jitter induced the XX decay and therefore significantly modifies the indistinguishability. On the other hand, the linewidth of the X transition keeps the same value as the one without stimulating lasers for all the delay times, confirming the fact that the improvement of the indistinguishability is due to the removal of the emission time jitter instead of other mechanisms, as presented in Fig.~4(b) and Extended Data Fig.~2. The linewidth of 5.42~GHz measured from a scanning Farby-Perot  interferometer is broader than the value of 2.1~GHz expected from indistinguishability of 0.926 because this QD suffers from spectral wandering in a long time scale ($\mu s$ to ms)\cite{PhysRevLett.116.213601}. Such a long time scale spectral wandering can be fully suppressed by stabilizing the charge environment with a PIN gate, as successfully demonstrated recently\cite{zhai2020low,tomm2021bright}. Essentially the linewidth is a slow measurement of indistinguishability due to the limited sampling rate of the scanning Fabry-Perot caivty while the Hong-Ou-Mandel interference characterizes the indistinguishability of photons separated by 13~ns. We note that the reduction of the indistinguishability could not be from the linewidth broadening due to other decoherence processes, e.g., pure dephasing and power broadening etc, otherwise the lower-bound of the linewidth would be 7.4~GHz for indistinguishability of 0.27. We observe that most of QDs in this wafer suffer from different extents of blinking behavior under TPE, which is probably due to the material impurities formed in the epitaxial growth~\cite{santori2001triggered}. In particular, the stimulating pulses suppress the blinking and improve the quantum efficiency (QE) from 0.3 to 0.39 for the investigated device (see details in Supplementary Information Section $\rm \uppercase\expandafter{\romannumeral6}$). The in-fiber efficiency of the source is calibrated as 0.03(2), which is limited by the transmission of the setup (0.096) and the QE of the QD (0.39). This number could be boosted up to 0.51(1) by optimizing the optics and the quality of QDs. E.g., using high-transmission optics with near-infrared coating can improve the transmission of the setup up to $\sim$0.6~\cite{ding2016demand} (see details in Supplementary Information Section $\rm \uppercase\expandafter{\romannumeral7}$) and a gated QD has exhibited a near-unity QE under TPE with a proper bias voltage~\cite{zhai2020low,schimpf2021entanglement}. Such a high in-fiber efficiency together with the ultra-high single-photon purity and high raw indistinguishability can immediately push this source to the state-of-the-art~\cite{wang2019towards,tomm2021bright,uppu2020scalable}.

\begin{center}
	\begin{figure*}
		\includegraphics[width=0.8\linewidth]{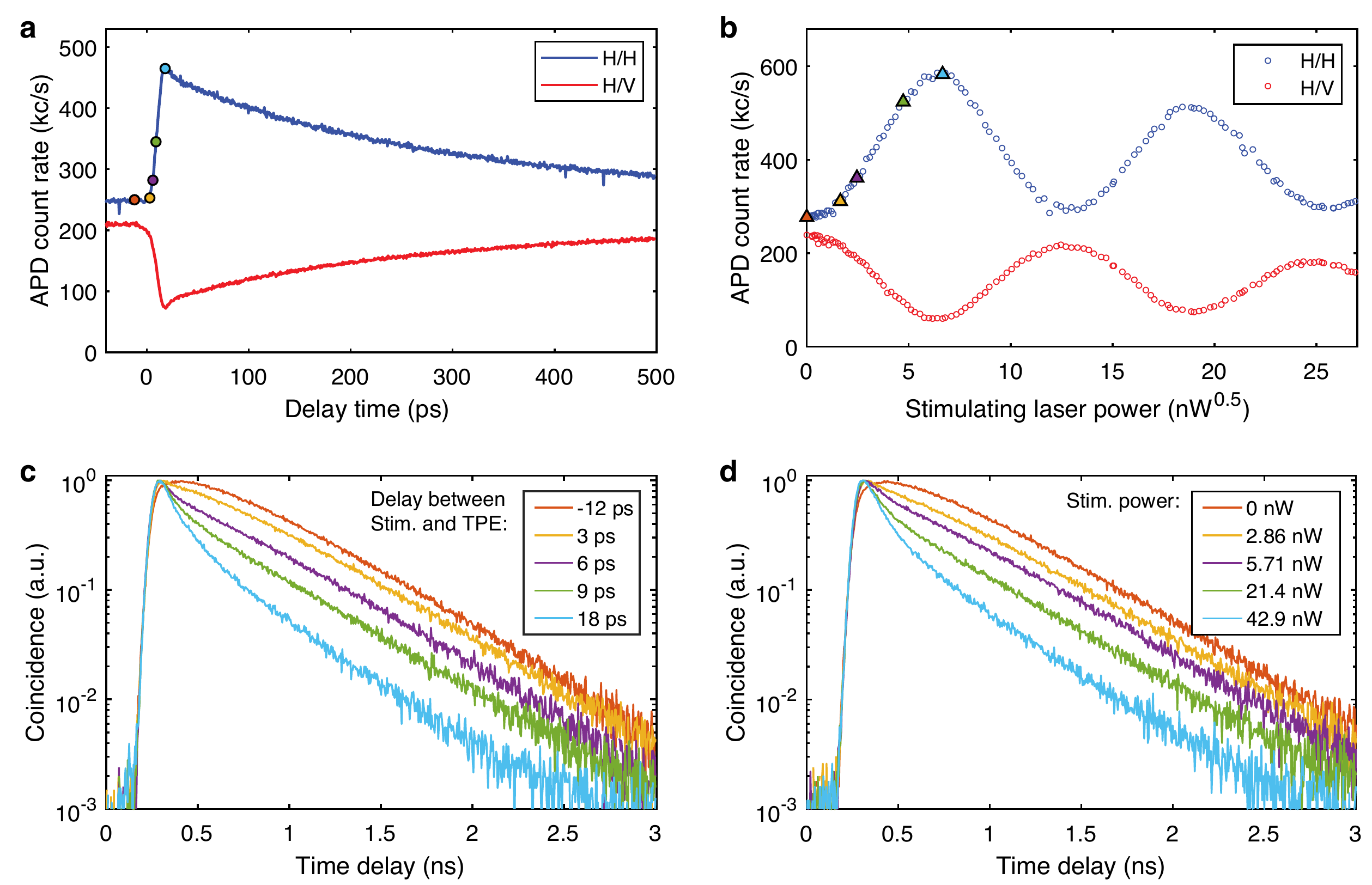}
		\caption{\textbf{Optimization of stimulation pulse parameters and control of the X decay dynamics.} (a) Intensities of the H- and V-polarized single photons emitted by X as a function of the delay time between the H-polarized stimulating pulses and TPE pulses. (b) Intensities of the H- and V-polarized single photons emitted by X as a function of power of the H-polarized stimulating pulses. (c) X decay dynamics with different delay times between the stimulating pluses and TPE pulses. The selected delay times are marked by circles in (a). (d) X decay dynamics with different powers of the stimulating pulses. The selected laser powers are marked by triangles in (b).}
		\label{fig:Fig5}
	\end{figure*}
\end{center}

\section{Optimization of stimulation pulse parameters and control of the exciton decay dynamics}

We further pinpoint the optimal parameters of the stimulating pulses used for boosting the performance of single-photon sources and present the opportunities of engineering the preparation and decay dynamics of X with the stimulating pulses. Fig.~5(a) presents the intensities of H- and V- polarized single photons as a function of the delay time between the H-polarized stimulating pulses and the TPE pulses. When the stimulating pulses arrive earlier than the TPE pluses, the H- and V-polarized single photons exhibit very similar intensities simply because there is no population in XX for the stimulating pulses to trigger the simulated emissions. The intensity of the H-polarized single photons increases rapidly once the delay time becomes positive and reaches to the maximal value of almost twice as the initial number at a delay time of 18 ps. Further increase of the delay time reduces the benefits from stimulating pulses since the XX has a high probability of randomly decaying to $\rm X_{H}$ or $\rm X_{V}$ via spontaneous emissions. For the V-polarized single photons, the intensity evolution follows the opposite trend of the H-polarized single photons due to the conservation of the XX population. With the optimal delay time of 18 ps, we plot the intensities of the H- and V-polarized single photons from X as a function of the stimulating laser power, as shown in Fig.~5(b). The oscillatory and complementary behaviors of $\rm X_{H}$ and $\rm X_{V}$ reveal the Rabi oscillation and the population conservation of the XX under the excitation of the stimulating pulses. We note that all the data shown in Fig.~2 and Fig.~3 are recorded under the optimal condition for the stimulating pulses (delay time of 18 ps and power of 40 nW). The delay time and the power of the stimulating pulses can also be exploited to engineer the preparation and decay dynamics of X, as presented in Fig.~5(c.d). By either adjusting the delay time or the laser power, the decay time of the X can be continuously tuned from 510~ps to 84~ps.

\section{Comparison of different excitation schemes}

Finally, we systematically compare the device performance as a polarized single-photon source under different excitation schemes. To avoid differences in performance due to variations of the epitaxial quality, we compare different excitation schemes through experiments performed with the same device. We implemented LA phonon-assisted excitation\cite{thomas2021bright}, resonant excitation, two-color resonant excitation\cite{he2019coherently,koong2021coherent} and the TPE with stimulating lasers on the same QD-micropillar device to exclusively compare the device metrics, as presented in Table I and Extended Data Fig.~3. The $g^{(2)}(0)$ under TPE with stimulating lasers and two-color resonant excitation are 0.002(1) and 0.004(1), much better than the results obtained under phonon-assisted excitation(0.030(1)) and resonant excitation(0.016(1)). Both resonant excitation and TPE with stimulating lasers exhibit very high-degree of photon indistinguishabilities of 0.940(3) and 0.926(4) respectively, significantly higher than those of LA-phonon assisted excitation (0.890(1)) and two-color resonant excitation (0.833(1)). One surprising observation is that the blinking behavior is absent (near-unity QE) in resonant excitation without any ancillary lasers or white lights while the QEs associated to the other excitation methods are rather poor even with the assistance of ancillary lasers or white lights. Such a blinking behavior under TPE excitation has been widely observed and it is attributed to the fluctuation of the charge environment\cite{doi:10.1021/acs.nanolett.7b00777}. A recent study has clearly demonstrated that the blinking behaviour under TPE can be fully suppressed and a near-unity QE is obtained by adding a suitable bias across the QD with a PIN gate \cite{schimpf2021entanglement}. Given the appreciable differences of QE, the in-fiber efficiency obtained under resonant excitation is slightly higher than those of TPE with stimulating lasers and two-color resonant excitation. The in-fiber efficiency of the LA-assisted excitation is much lower probably due to the low exciton preparation rate in the phonon-assisted process (the exciton preparation rate is not included in definition of the QE associated with the blinking behavior). By optimizing the setup with high transmission opitcs and implementing PIN gates, the achievable in-fiber efficiency of TPE with stimulating lasers is 0.510(1), twice of the achievable in-fiber efficiency of 0.251(1) under resonant excitation. Therefore, our method could potentially deliver a polarized single photon source with two-time brightness, much better single-photon purity and similar photon indistinguishability compared to the resonant excitation scheme.

\begin{table*}[ht!]
	\centering
	\vspace{0.1cm}
	\begin{threeparttable}
	\caption{Comparison of the device performances of the same devices under different excitation schemes. The metrics are extracted from Extended Data Fig.~3}
	\label{Table1}
	\begin{tabular}{|c|c|c|c|c|}
		\hline
		\textbf{Metrics} \vspace*{\fill}  &   \textbf{LA-phonon assisted excitation}  \vspace*{\fill} &    \textbf{Resonant excitation} &  \textbf{Two-color resonant excitation.}    &  \textbf{TPE-Stim.} \\
		\hhline{|-|-|-|-|-|}
		\textbf{$g^{(2)}(0)$} & 0.030(1) & 0.016(1) & 0.004(1) & 0.002(1)   \\
		\hline
		\textbf{HOM} & 0.890(1) & 0.940(3) & 0.833(1) & 0.926(4) \\
		\hline
		\textbf{QE} & 0.47(1) & 0.98(2) & 0.39(1) & 0.39(1)  \\
		\hline
		\textbf{in-fiber efficiency} & 0.007(1) & 0.037(1) & 0.023 (1) & 0.030(2) \\
		\hline
		\textbf{in-fiber efficiency with optimizations} & 0.099(1) & 0.251(1) & 0.391(1) & 0.510(1) \\
		\hline
		\end{tabular}
	\end{threeparttable}
\end{table*}

\section{Conclusions}

To conclude, we exploit stimulating pulses with the same energy as XX of a QD to tailor the spontaneous emission of X. Our method doubles the source brightness achievable in resonance fluorescence mode under the cross-polarization configuration. The fundamental limit of the indistinguishability under TPE is overcome by removing the time jitter associated to the spontaneous emission of XX via the stimulated emissions. This approach can be, in principle, immediately applied to the existing single-photon sources based on QDs, e.g., QDs in nanowires~\cite{,claudon2010highly}, microlens\cite{gschrey2015highly} and photonic crystal waveguides~\cite{uppu2020scalable}, for further boosting the device performance. Moving forwards, more advanced photonic structures and high-quality QD wafer could be employed to pursue the ultimate single-photon sources with our approach. For instance, triply resonant cavities~\cite{marty2021photonic} could be used to simultaneously enhance the TPE efficiency, stimulated emission of XX and spontaneous emission of X. Gated QD devices can suppress the charge noise to ensure the unity QE of QDs. Our work provides exciting opportunities of exploiting quantum emitters beyond a two-level system for exploring non-Hermitian physics in quantum regime\cite{naghiloo2019quantum} and developing novel quantum nanophotonic devices such as quantum logic gates\cite{li2003all}, single-photon nonlinearities between two optical beams~\cite{nguyen2018giant} and nanolasers with single exciton gain~\cite{park2021colloidal}.

\textit{Note added}: During the peer review process, we became aware of a work describing a similar methodology for bulk QDs\cite{sbresny2021stimulated}.

\noindent \textbf{Acknowledgements}
J.~L. thanks Kartik Srinivasan for his continuing support and Hen Shen for helpful discussions. This research was supported by National Key R\&D Program of China (2018YFA0306101), Key-Area Research and Development Program of Guangdong Province (2018B030329001), the National Natural Science Foundation of China (62035017, 11874437, 12074442, 91836303) the Local Innovative and Research Teams Project of Guangdong Pearl River Talents Program (2017BT01X121) and the National Super-Computer Center in Guangzhou.

\noindent \noindent \textbf{Author Contributions}
J.~L conceived the project. S.~F.~L performed to the numerical simulations. Y.~Y, X.~B.~S, S.~L.~L, X.~J.~S, H.~Q.~L, H.~M.~H, H.~Q.~N grew the quantum dot wafers. S.~F.~L and X.~S.~L fabricated the devices. J.~I~S contributed to the theoretical modeling. Y.~M.~W, S.~F.~L and J.~L built the setup and characterized the devices. J.~L, S.~F.~L and Y.~M.~W analyzed the data. J.~L wrote the manuscript with inputs from all authors. J.~L, S.~Y.~Y, Z.~C.~N. and X.~H.~W supervised the project.

\noindent \noindent \textbf{Conflict of Interest:}
The authors declare no conflict of interest.


\section{methods}

\noindent \textbf{Sample fabrication:}
We use a III-V wafer consists a single layer of low density In(Ga)As QDs between 18(28) top(bottom)  GaAs/$\rm{Al_{0.9}Ga_{0.1}As}$ distributed Bragg reflector (DBRs) grown via molecular beam epitaxy. The deterministically coupled QD-micropillar is fabricated by using the fluorescence imaging technique. First, metallic alignment marks are created by using E-beam lithography, metal evaporation and lift-off processes. Then the wide-field PL-images containing both alignment marks and the QDs are obtained by a bi-chromatic illumination technique. The positions of the QDs respective to the alignment marks are extracted from the fluorescence images. Finally, micropillars are realized through the second E-beam lithography followed by a chlorine-based dry etch process. The diameter of micropillar is carefully chosen to ensure the spectral resonance between the X state of the QD and the fundamental cavity mode of the micropillar, The detailed fabrication flow of the deterministically coupled micropillar-QD devices is presented at Supplementary Fig.~S1.

\noindent \textbf{Optical characterizations:}
The schematic of the setup for optical characterizations is presented in Extended Data Fig.~1. The sample is located in a closed-circle cryostat with a base temperature of 5.1 K. A femto-second pulse from a Ti-sapphire oscillator is split and shaped into 13~ps pulses with different colors for TPE and stimulating excitation. A delay line is used to vary the delay time between the TPE pulse and the stimulating pulse. The TPE pulse and stimulating pulse are combined in a single-mode fiber and sent into a customized confocal microscope with polarization controllers in both the excitation and collection paths. A white light lamp in cryogenic confocal system is used as an illumination source for imaging and meanwhile reduces the charge noise around the QD for improving the quantum efficiency (QE) of the QD under TPE. The emitted single photons are collected by a single-mode fiber and sent to a spectrometer for spectral analysis. A filter-set consisting of a volume grating and notch filters is employed to spectrally separate the single photons from the TPE laser and simulated pulses before the HBT and HOM interferometers. The delay time in the HOM interferometer is chosen to be 12.56 ns to match the repetition rate of the excitation laser. The linewith is measured with a scanning Farby-Perot interferometer with resolution of 94~MHz.

\noindent \textbf{Data availability:}
The data that support the findings of this study are available within the paper and the Supplementary Information. Source data are provided with this paper. Other relevant data are available from the corresponding authors on reasonable requests.

\newpage
\onecolumngrid \bigskip

\begin{center} {{\bf \large EXTENDED DATA}}\end{center}

\setcounter{figure}{0}
\makeatletter
\renewcommand{\thefigure}{E\@arabic\c@figure}

\begin{center}
	\begin{figure}[!h]
		\includegraphics[width=1\linewidth]{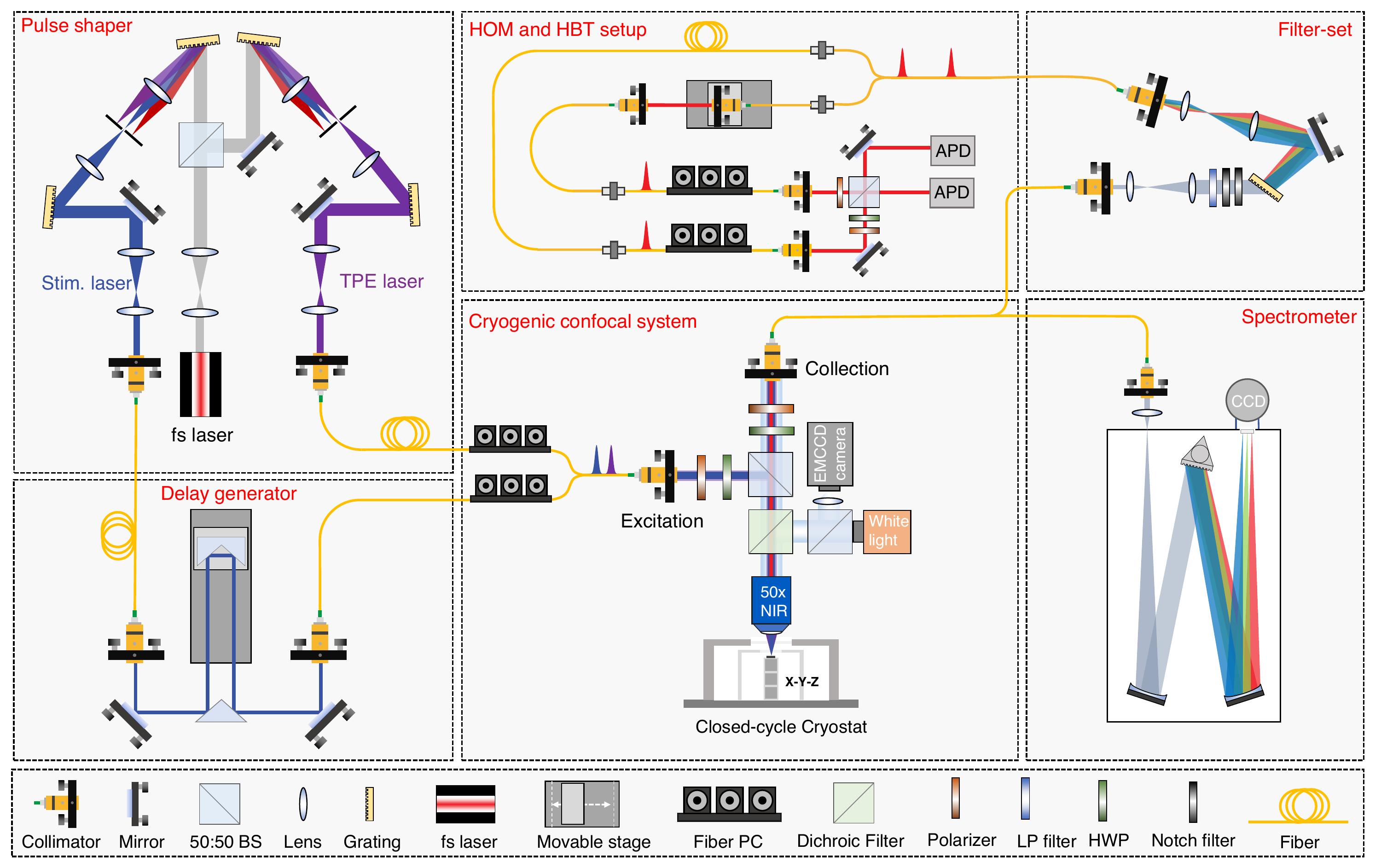}
		\caption{\textbf{Schematic of the setup for optical characterizations.} The setup consisting of 6 functional sections including cryogenic confocal system, pulse shaper, delay generator, spectrometer, filter set and HOM/HBT interferometers. BS: beam splitter, HWP: half-wave plate, LP filter: Long pass filter.}
		\label{fig:Fig5}
	\end{figure}
\end{center}

\newpage

\begin{center}
	\begin{figure}[!h]
		\includegraphics[width=1\linewidth]{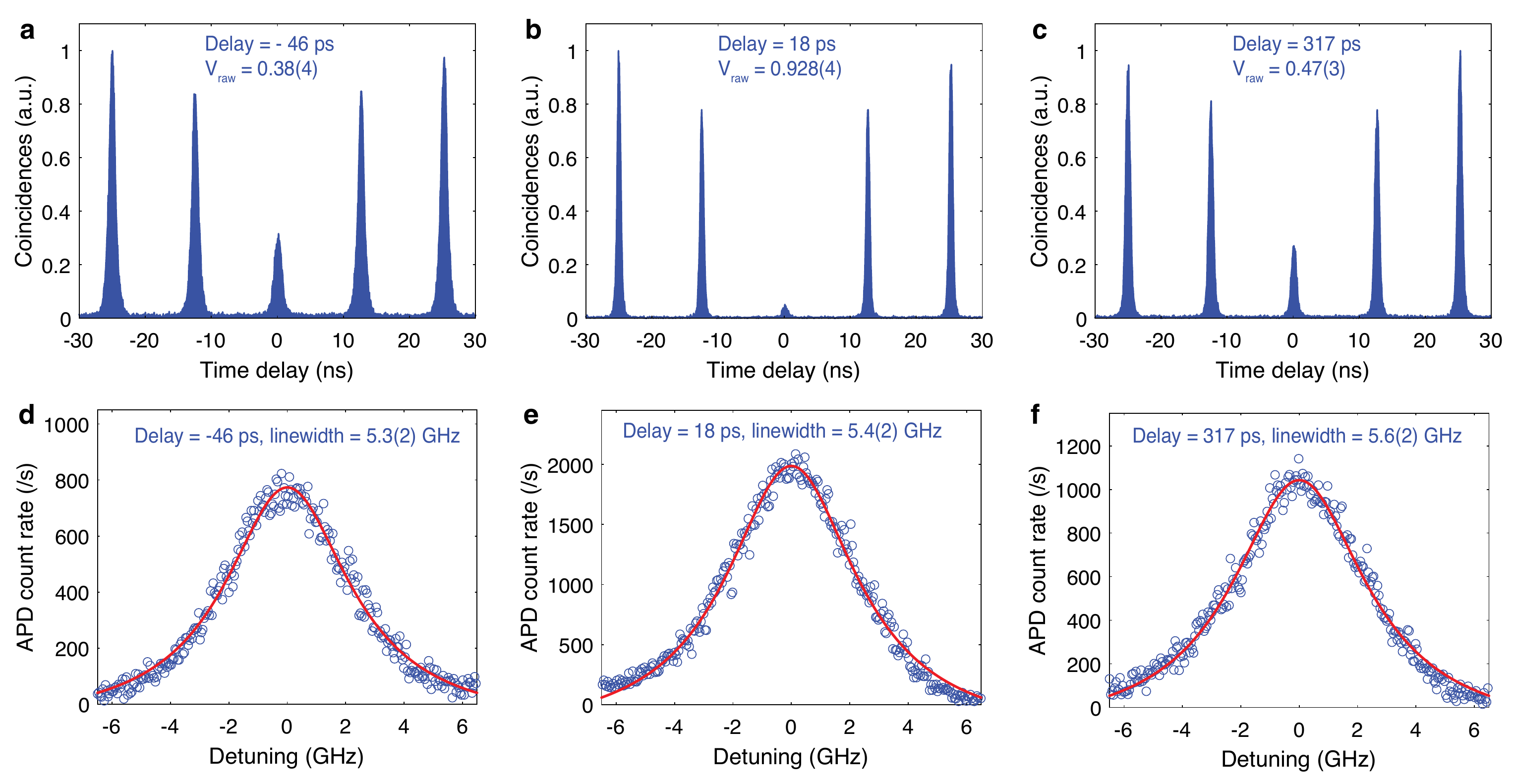}
		\caption{\textbf{Indistinguishability and linewidth for different delay time.} The raw visibilities of the HOM interference for delay time of – 46 ps (a), 18 ps (b) and 317 ps (c), respectively. The linewidths of the emission for delay time of -46 ps (d), 18 ps (e) and 317 ps (f), respectively.}
		\label{fig:Fig5}
	\end{figure}
\end{center}

\newpage

\begin{center}
	\begin{figure}[!h]
		\includegraphics[width=1\linewidth]{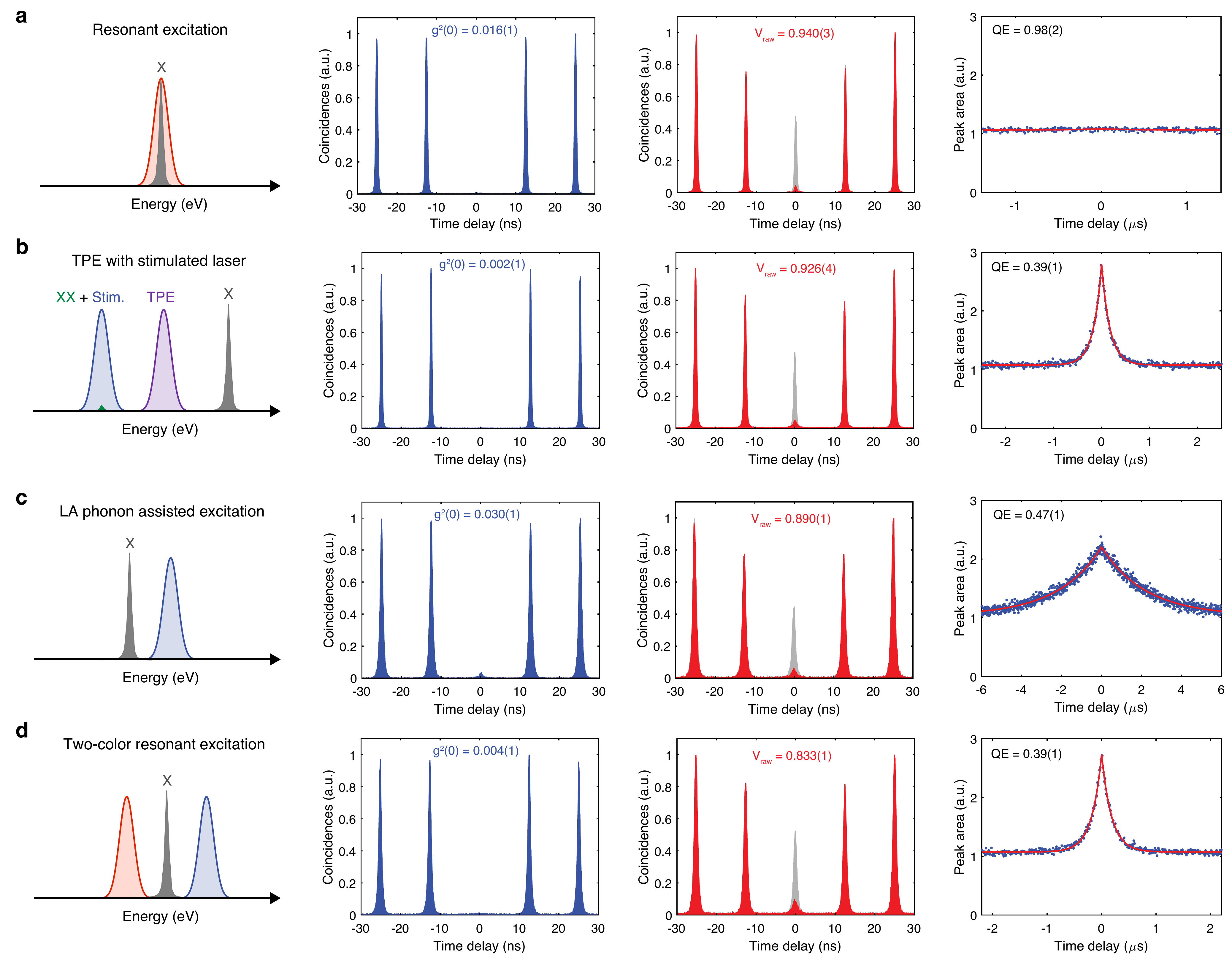}
		\caption{\textbf{Comparison of the metrics for the source under different excitation schemes.}  Excitation scheme, second-order correlation, Hong-Ou-Mandel interference and Blinking behavior for resonant excitation (a), TPE with stimulating lasers (b), LA-phonon assisted excitation (c) and two-color resonant excitation (d).}
		\label{fig:Fig5}
	\end{figure}
\end{center}

\newpage
\onecolumngrid \bigskip

\begin{center} {{\bf \large SUPPLEMENTARY INFORMATION}}\end{center}

\setcounter{figure}{0}
\setcounter{section}{0}
\makeatletter
\renewcommand{\thefigure}{S\@arabic\c@figure}

\section{Sample fabrication}

We use a III-V wafer consists a single layer of low density In(Ga)As QDs between 18(28) top(bottom)  GaAs/$\rm{Al_{0.9}Ga_{0.1}As}$ distributed Bragg reflector (DBRs) grown via molecular beam epitaxy. The deterministically coupled QD-micropillar is fabricated by using the fluorescence imaging technique~\cite{liu2017cryogenic}. First, metallic alignment marks are created by using E-beam lithography, metal evaporation and lift-off processes. Then the wide-field PL-images containing both alignment marks and the QDs are obtained by a bi-chromatic illumination technique~\cite{liu2017cryogenic}. The positions of the QDs respective to the alignment marks are extracted from the fluorescence images. Finally, micropillars are realized through the second E-beam lithography followed by a chlorine-based dry etch process. The diameter of micropillar is carefully chosen to ensure the spectral resonance between the X state of the QD and the fundamental cavity mode of the micropillar.

\begin{figure}[!h]
	\includegraphics[width=0.9\linewidth]{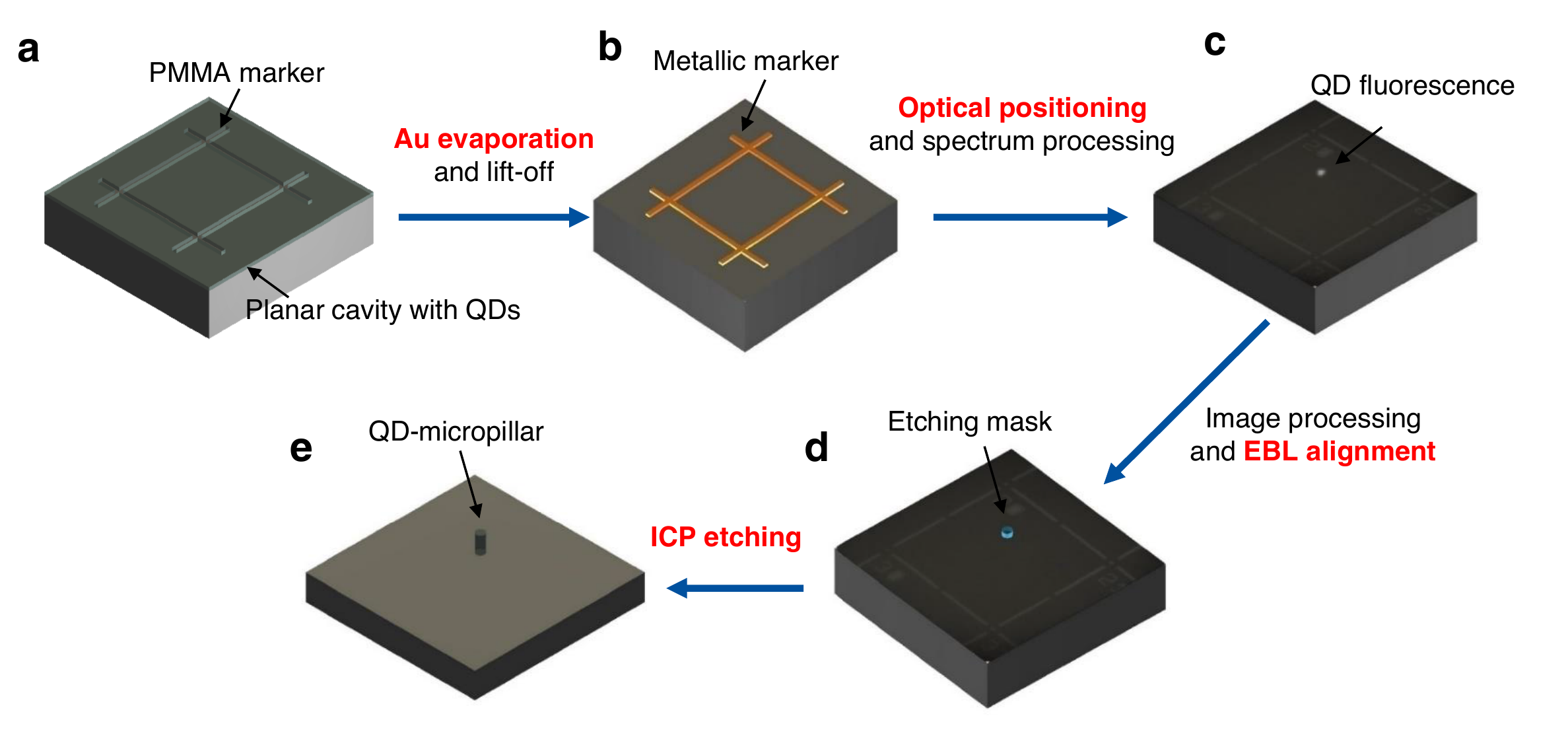}
	\caption{Fabrication flow of deterministically coupled QD-micropillar devices.}
	\label{SIfig:Fig1}
\end{figure}

\newpage
\section{Sample information}

A scanning electron beam (SEM) image of the fabricated micropillar with a diameter of 2.5 $\rm{\mu}$m is shown in Fig.~S2(a). The micropillar consist of a $\lambda$-GaAs cavity sandwiched between 18 pairs of top DBRs and 28 pairs of bottom DBRs. The QD layer is in the middle of the $\lambda$ cavity. The propagation and far-field pattern of the cavity mode presented in Fig.~S2(b, c) are simulated with 3D-FDTD method. Most of the photons in the cavity mode are emitted upwards and the Gaussian-like far-field mode profile enables a very  high collection efficiency by an objective or single-mode fiber. The divergent angle is below 30$^\circ$ so that 99\% of the upward emitted photons can be collected by an objective with N.A. of 0.65, as we used in this experiment.

\begin{figure}[!h]
	\includegraphics[width=0.9\linewidth]{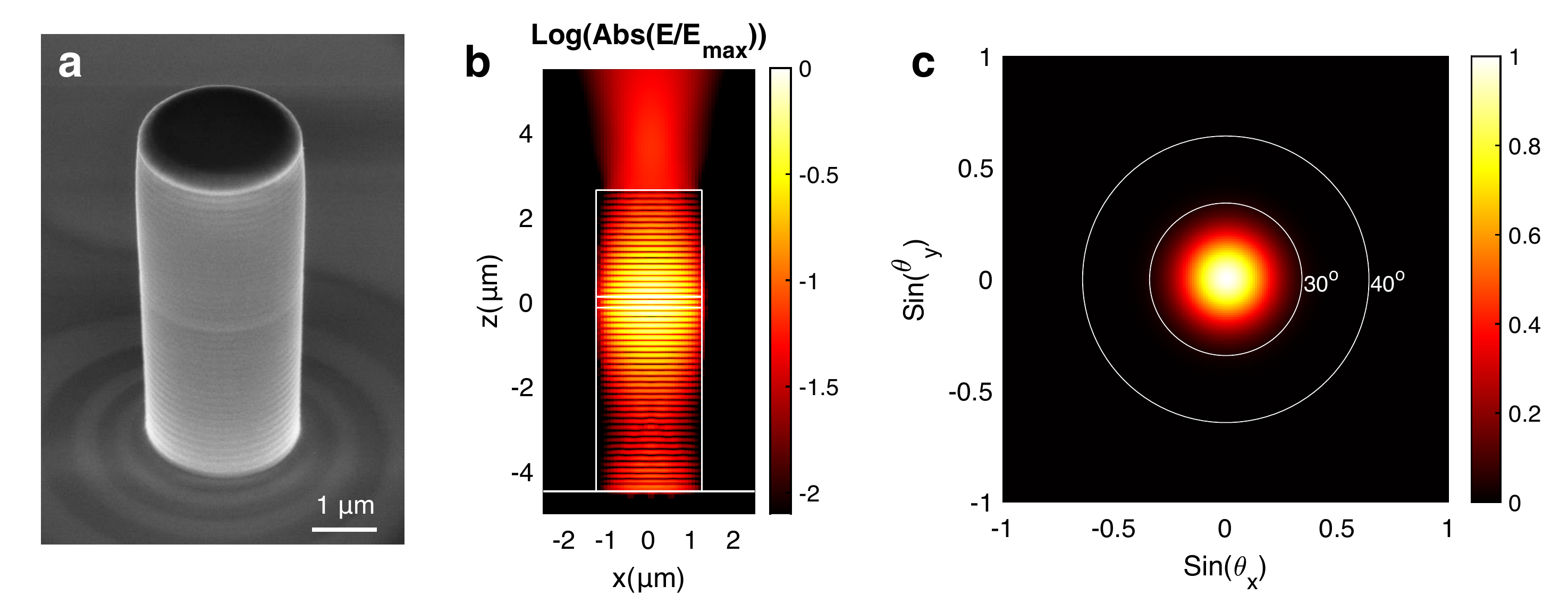}
	\caption{Sample information. (a) SEM image of the fabricated micropillar cavity. (b) The emission profile of the cavity mode, showing most of the photons are emitted upwards. (c) Far-field pattern of the cavity mode, exhibiting a Gaussian-like profile with a divergent angle below 30$^\circ$.}
	\label{SIfig:Fig2}
\end{figure}

\newpage
\section{Modelling the indistinguishability}

To simulate the experiment, we consider a model of a four level system consisting of the biexciton ($\ket{XX}$), horizontal ($\ket{X_H}$) and vertical ($\ket{X_V}$) polarized excitons, and the unoccupied ground state $\ket{0}$, as shown in Fig.~S3(a).
The free Hamiltonian for the above model takes the form:
\begin{equation}
	H_S = E_{XX} \ket{XX}\!\bra{XX} + E_{XH} \ket{X_H}\!\bra{X_H} + E_{XH} \ket{X_V}\!\bra{X_V}, 
\end{equation}
where $E_{XX} = 2 E_0$ is the biexciton energy, $E_{XH} = E_0 + \Delta F$ is the energy of the horizontally polarized exciton, and $E_{XV}= E_0 - \Delta F$ is the vertical polarized state. Here we have introduced the exciton splitting as $E_0$, and $\Delta F$ is the fine structure splitting. 

\begin{figure}[h]
	\centering
	\includegraphics[width=0.8\linewidth]{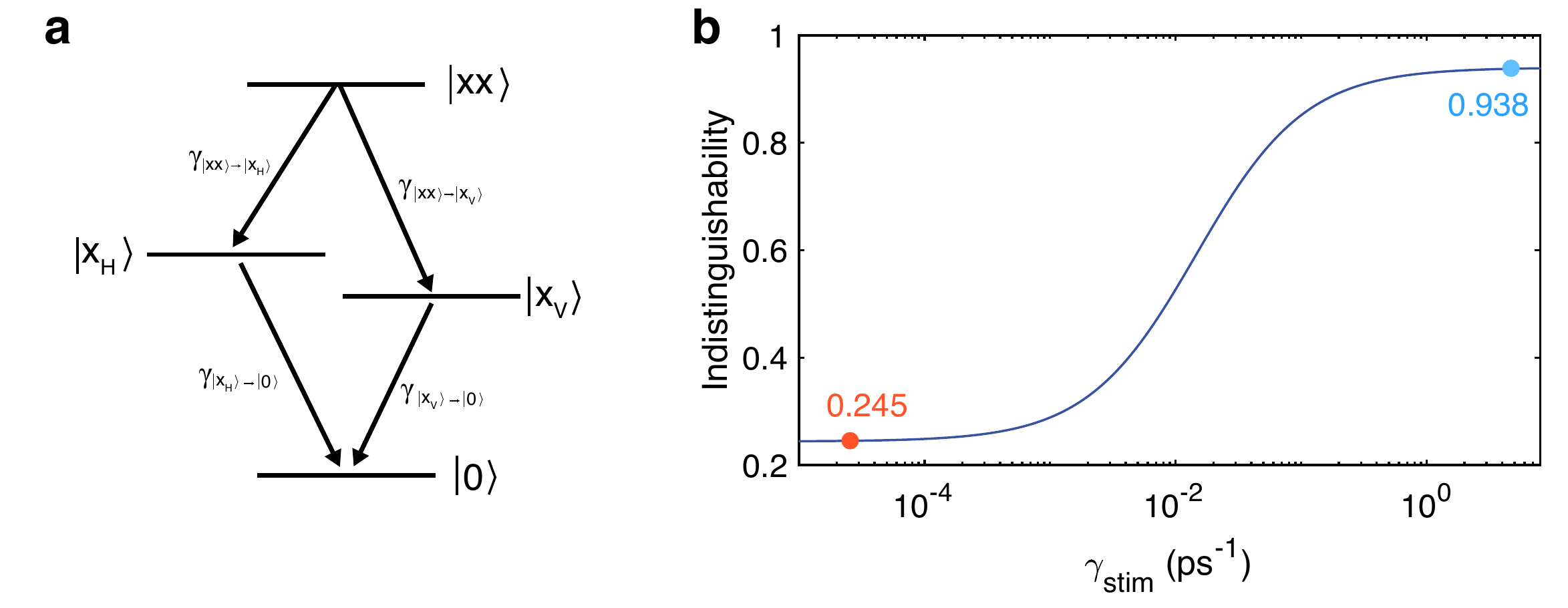}
	\caption{(a) The level diagram for the  associated transition rates. (b) Indistinguishability as a function of the stimulated rate.}
	\label{fig:my_label}
\end{figure}

We can describe the above model phenomenologically using the Lindblad form master equation:
\begin{equation}
	\partial_t \rho(t)= -i\left[H_S,\rho(t)\right] + \sum_{\alpha = XV, XH} \left(\frac{\gamma_{XX\rightarrow \alpha}}{2}\mathcal{L}(\ket{XX}\!\bra{\alpha})
	+\frac{\gamma_{\alpha\rightarrow 0}}{2}\mathcal{L}(\ket{\alpha}\!\bra{0})\right)
	+\sum_{\alpha=XX,XV,XH}\frac{\Gamma}{2}\mathcal{L}(\ket{\alpha}\!\bra{\alpha})
\end{equation}
where we have introduced the density operator $\rho(t)$ for the 4-level system, and the Lindblad superoperator $\mathcal{L}(O) = 2O^\dagger\rho(t) O - O^\dagger O\rho(t) - \rho(t)O^\dagger O$. 
Notice that the $\alpha$ index determines which of the horizontal and vertical polarization pathways the dissipator is associated with.
The first sum over Lindblad terms correspond to optical transitions, while they second are pure dephasing processes of the energy levels which we assume occur at the same rate $\Gamma$.

To account for the stimulated emission process, we consider the decay rate from the biexciton to the horizontal branch to take the form $\gamma_{XX\rightarrow XH}\rightarrow \gamma_{XX\rightarrow XH}+ \gamma_\mathrm{stim}$.
The first term corresponds to the decay from the biexciton state in the absence of driving which we choose, while $\gamma_\mathrm{stim}$ accounts for the stimulated emission.
In addition to the stimulation of the $\ket{XX}\rightarrow\ket{XH}$ transition, the $\ket{XH}\rightarrow\ket{0}$ is enhanced through a cavity, such that $\gamma_{XH\rightarrow0}= F_P\gamma_X$, where $F_P$ is the Purcell factor of the cavity and $\gamma_X$ is the bare exciton decay rate. We assume that the bare decay rates of both single exciton states are the same, such that, $\gamma_{XV\rightarrow0}=\gamma_X$.

By initializing our system in the biexciton state $\rho(t=0) = \ket{XX}\!\bra{XX}$, we can gain an analytic expression for the above master equation, yielding the populations:
\begin{align}
	P_{XX}(t) =& e^{-(\gamma_{XX\rightarrow XV}+\gamma_{XX\rightarrow XH} + \gamma_\mathrm{stim})t},\\
	P_{XH}(t) =& \frac{\gamma_{XX\rightarrow XH} + \gamma_\mathrm{stim}}{F_P\gamma_{X} - \gamma_{XX\rightarrow XV}-\gamma_{XX\rightarrow XH}  - \gamma_\mathrm{stim}}\left(e^{-(\gamma_{XX\rightarrow XV}+\gamma_{XX\rightarrow XH} + \gamma_\mathrm{stim})t} - e^{-F_P\gamma_{X}t}\right),\\
	P_{XV}(t) =& \frac{\gamma_{XX\rightarrow XV}}{\gamma_X-\gamma_{XX\rightarrow XV}-\gamma_{XX\rightarrow XH} - \gamma_\mathrm{stim}}\left(e^{-\gamma_X t}-e^{-( \gamma_\mathrm{stim}+\gamma_{XX\rightarrow XH}+\gamma_{XX\rightarrow XV})t}\right).
\end{align}

To estimate the indistinguishability of photons emitted through the $XH\rightarrow0$ transition, we require the first-order correlation function for this transition, that is $g^{(1)}_{XH\rightarrow0}(t,\tau) = \langle\sigma^\dagger(t+\tau)\sigma(t)\rangle$ with $\sigma=\ket{g}\!\bra{XH}$.
This correlation function can be calculated through the quantum regression theorem, such that:
\begin{equation}
	g^{(1)}_{XH\rightarrow0}(t,\tau) =\frac{(\gamma_{XX\rightarrow XH} + \gamma_\mathrm{stim})
		e^{(iE_{XH}-\Gamma/2)\tau} e^{-F_P\gamma_X\tau/2}
	}{F_P\gamma_{X} - \gamma_{XX\rightarrow XV}-\gamma_{XX\rightarrow XH}  - \gamma_\mathrm{stim}} (e^{-(\gamma_{XX\rightarrow XV}+\gamma_{XX\rightarrow XH} +\gamma_\mathrm{stim})t} -e^{-F_P\gamma_X t}).
\end{equation}
From this, we can calculate an expression for the indistinguishability using the standard expression\cite{kaer2013role}:
\begin{equation}\begin{split}
		\mathcal{I} = \frac{\int_0^\infty dt\int_0^\infty d\tau \left\vert g^{(1)}_{XH\rightarrow0}(t,\tau)\right\vert^2}{\int_0^\infty dt\int_0^\infty d\tau P_{XH}(t)P_{XH}(t+\tau)} 
		= \frac{F_P\gamma_X}{\Gamma + F_P\gamma_X}\frac{\gamma_{XX\rightarrow XH}+\gamma_{XX\rightarrow XV}+ \gamma_\mathrm{stim}}{ \gamma_{XX\rightarrow XH}+\gamma_{XX\rightarrow XV}+F_P\gamma_X + \gamma_\mathrm{stim}}.
\end{split}\end{equation}
The expression derived above shows has two key contributions, the first factor is identical to the indistinguishability for the resonant excitation regime, the second is associated to the time-jitter. In the limit that the stimulated process dominates, such that $\gamma_\mathrm{stim}\rightarrow\infty$, the indistinguishability tends to the value obtained through resonant excitation. This is a consequence of the stimulated emission process effectively removing the effect of timing jitter. The above expression is consistent with those derived in Ref.~\onlinecite{unsleber2015two} for a three level system.

To calculate the the indistinguishability, we take the bare exciton lifetime of $\gamma_X^{-1} = 611$~ps, Purcell factor of $F_P=6.6$, and assume that the biexciton decay rates are equal such that $\gamma_{XX}^{-1}= \gamma_{XX\rightarrow XH}^{-1}=\gamma_{XX\rightarrow XV}^{-1}=528$~ps. To estimate the pure dephasing rate we use the indistinguishability measured through strict resonant excitation, i.e. $\mathcal{I}_\mathrm{res} = F_P\gamma_X/(F_P\gamma_X +\Gamma)=0.94$, this yields $\Gamma^{-1}=1450$~ps.

With these values we can calculate the indistinguishability as a function of the stimulated emission rate $\gamma_\mathrm{stim}$, as shown in Fig.~S3(b). We obtain an indistinguishability of $0.94$ in the limit $\gamma_{stim}\rightarrow\infty$ and in the presence of pure dephasing, this is consistent with the measured value of 0.926. In the absence of the stimulated process, the above model predicts the indistinguishability of 0.245, which agrees very well with our experiment (0.27$\pm0.03$).

\newpage
\section{Extraction of the Purcell factor}
The Purcell factor is extracted by measuring the lifetimes of the QD resonant with and far-detuned from the cavity mode under resonant excitation condition, as shown in Fig.~S4. The on-resonance lifetime ($\tau_{on}$) of 80~ps and the far-detuned lifetime ($\tau_{off}$) of 611.6~ps give rise to a Purcell factor $F{_p}=\frac{\tau_{off}}{\tau_{on}}-1=6.6$\cite{unsleber2016highly}. The oscillation in the decay curve for far-detuned condition is due to the beating between the two excitons with a fine-structure splitting.\cite{scholl2019resonance,ollivier2020reproducibility}
\begin{figure}[!h]
	\includegraphics[width=0.8\linewidth]{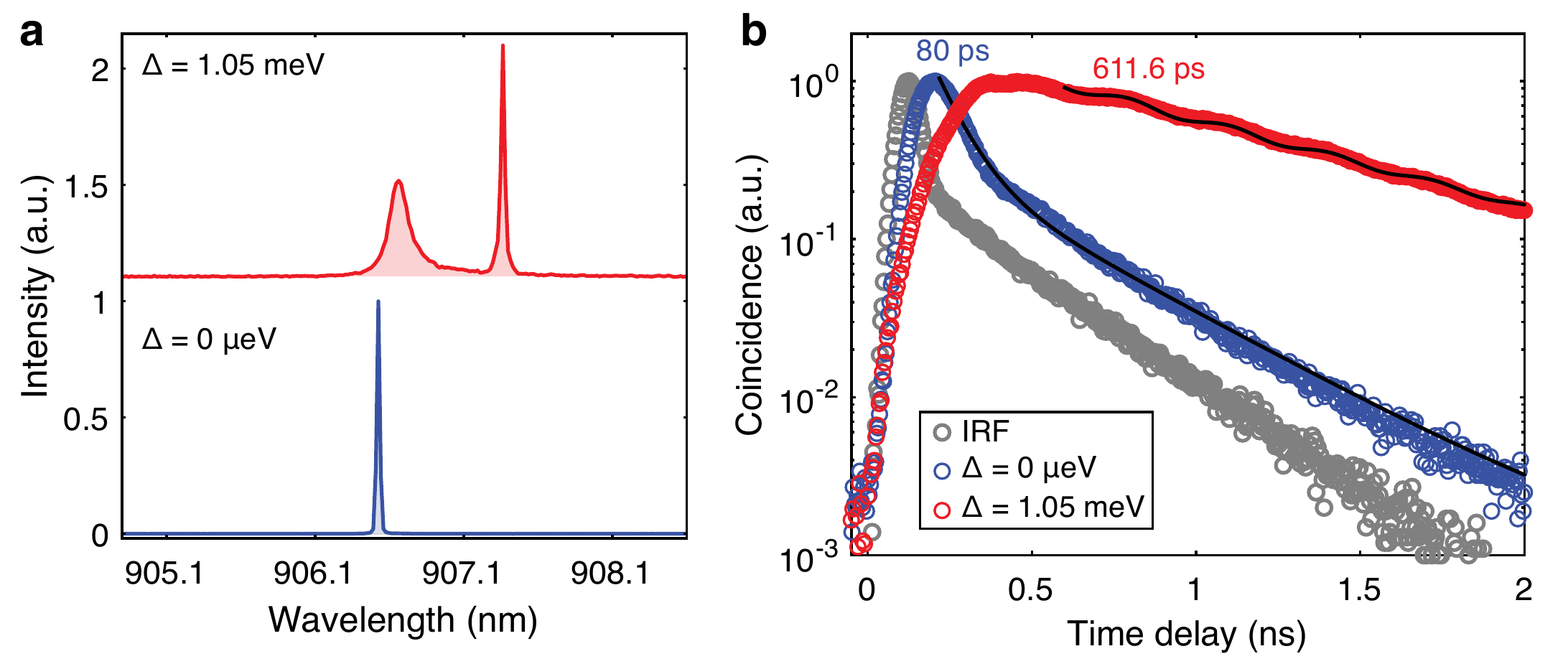}
	\caption{Extraction of the Purcell factor. (a) Resonant excitation  spectra of the QD-micropillar system with a detuning of 0~meV and 1.05~meV. (b) Lifetime measurements of QD resonant with and far-detuned from the cavity mode under resonant excitation condition.}
	\label{SIfig:Fig5}
\end{figure}

\newpage
\section{Polarization characteristics of the QD and the micropllar}
The polarization-dependence of the energy for the X is plotted in Fig.~S5(a), showing a fine structure splitting (FSS) of 14.2 $\mu$eV. The cavity presents two orthogonal modes, with a mode splitting of 32.6 $\mu$eV caused by small geometric anisotropy induced during fabrication process, as shown in Fig.~S5(b). The quality (Q) factor of each mode labeled by H-cavity and V-cavity are extracted as 7539.5 and 6958.5, respectively. The angle between the $\rm{X_H}$ and the H-polarized cavity mode is 37.5$^\circ$ as presented in Fig.~S5(c). A high-resolution spectrum (Fig.~S5(d)) of the QD and cavity mode indicates that X couples to the H-polarized cavity mode slightly better than the V-polarized cavity mode, which results in weakly H-polarized single photon emission under TPE excitation, as shown in Fig.~3(a) of the maintext.

\begin{figure}[!h]
	\includegraphics[width=0.9\linewidth]{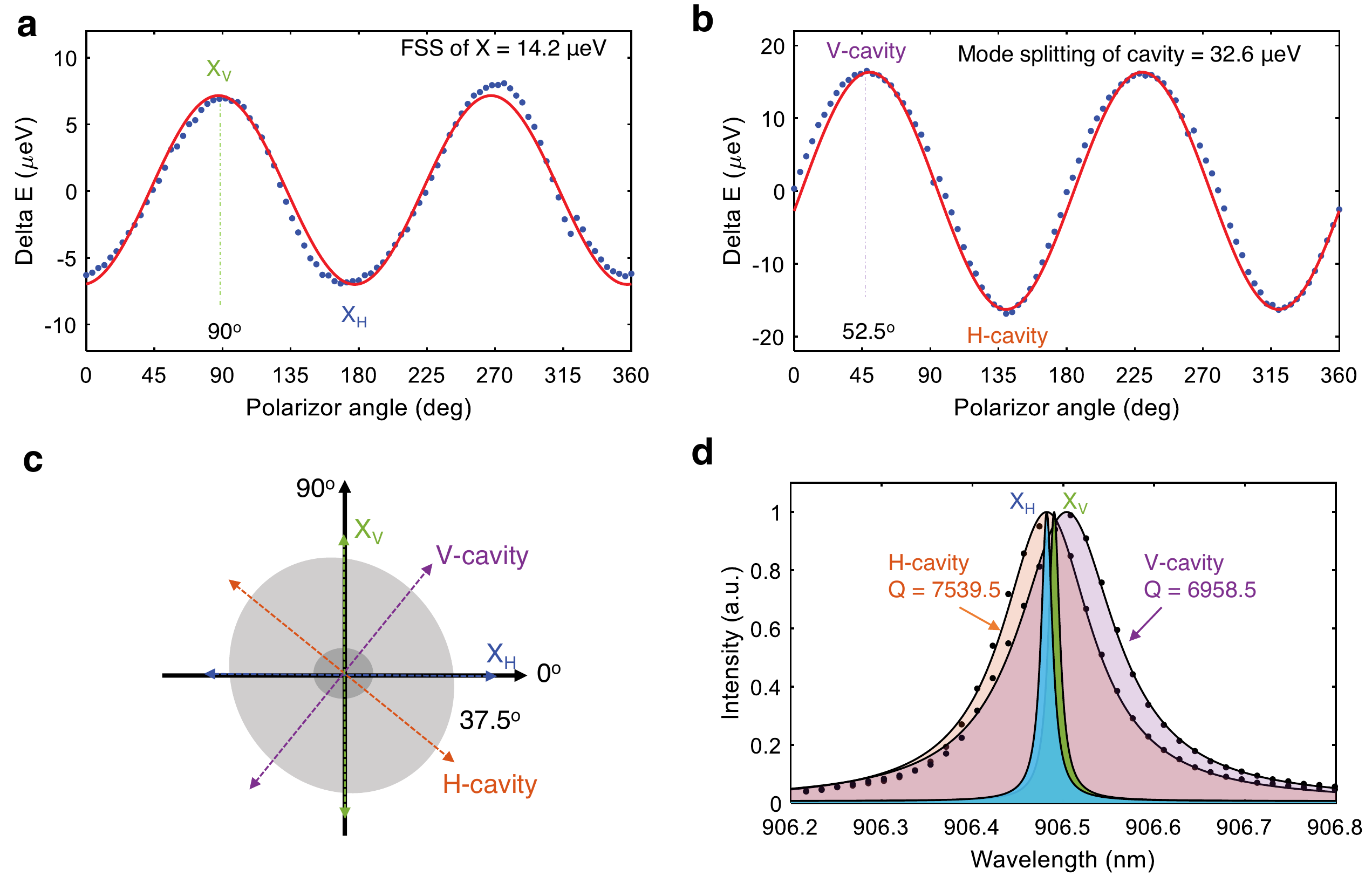}
	\caption{Polarization characteristics of the QD and the micropillar. (a) Polarization dependence of the energy of the X. (b) Polarization dependence of the energy of the cavity mode. (c). The angle of the polarization axis between the X and cavity mode. (d). High-resolution spectrum of the X and the cavity modes. }
	\label{SIfig:Fig6}
\end{figure}

\newpage
\section{Quantum efficiencies (QE) of the QD under different excitation scheme}
The investigated QD exhibit different extents of the blinking behavior under different excitation schemes, as shown in Fig.~S6. The QEs of the QDs can be extracted by fitting the long-time scale $g^{(2)}$ function with the following model~\cite{santori2001triggered}:

\begin{equation}\label{key}
	\rm I(t) = 1+\frac{\tau_{off}}{\tau_{on}}\times exp\left(-(\frac{1}{\tau_{off}}+\frac{1}{\tau_{on}})\times |t|\right)
\end{equation}
where I(t) is the normalized correlation peak area of time delay t, $\rm \tau_{on}$ and $\rm \tau_{off}$ reprent the time constants of "bright" and "dark" state of the QD. the QE is caculated with:
\begin{equation}\label{key}
	\rm{QE} = \frac{\tau_{on}}{\tau_{on}+\tau_{off}}
\end{equation}
Under the TPE, the QE is as low as 0.057. Adding a very weak while light illumination improves the QE to 0.3. By further introducing the stimulating pulses, we obtain the QE of 0.39.

\begin{figure}[!h]
	\includegraphics[width=0.9\linewidth]{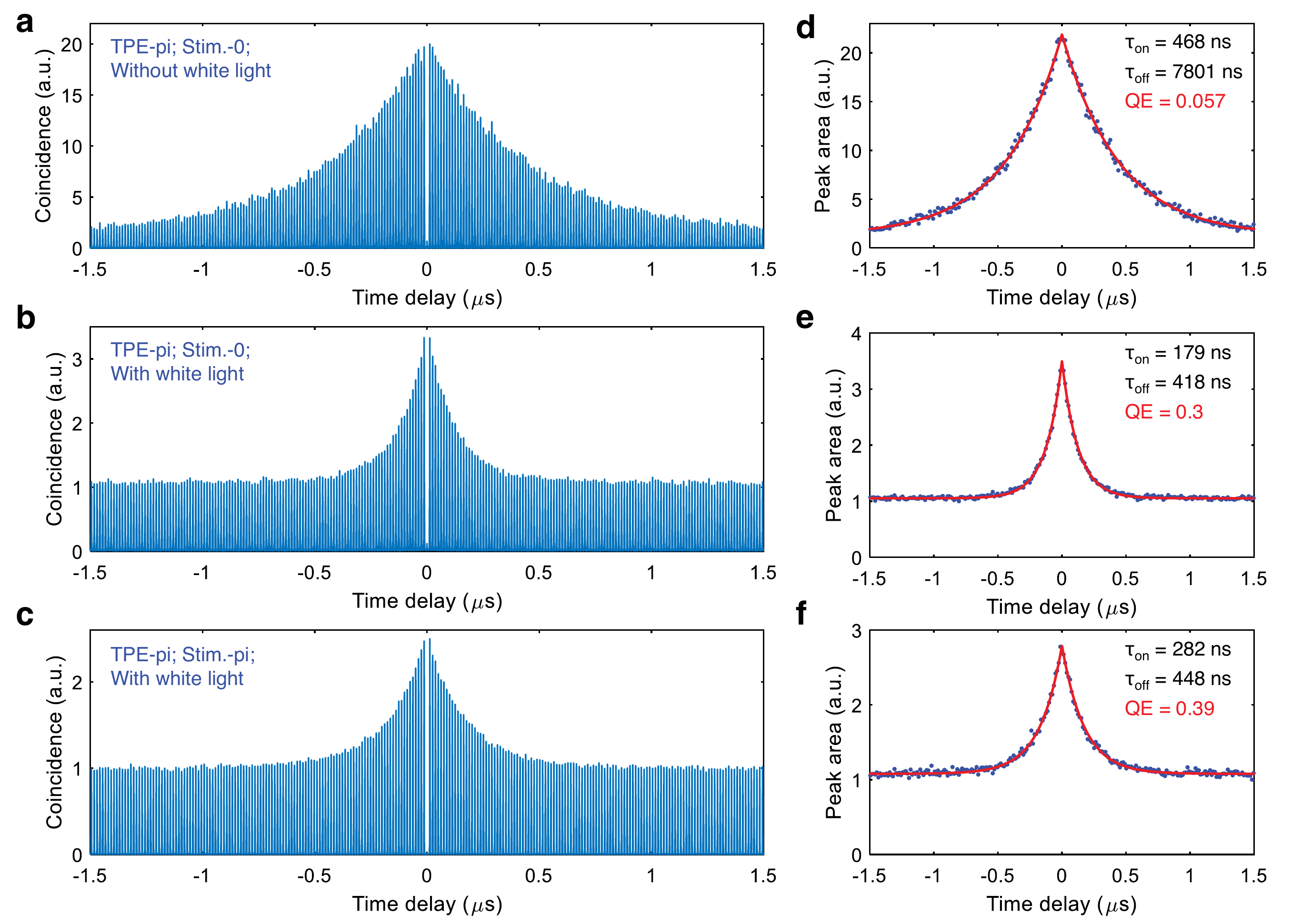}
	\caption{Blinking behavior of the investigated QD under different excitation schemes. (a), (b), (c) are the histograms of $g^{(2)}$ under TPE, TPE with white light and TPE with white light and stimulating pulses. (d), (e), (f) are the peak areas and the fits for the coincidence events in (a), (b) and (c).}
	\label{SIfig:Fig7}
\end{figure}

\newpage
\section{Source brightness}
We obtained 0.602 and 0.316 million counts/s in the in Si-based avalanche photon diode (APD) with and without the stimulating pulses. By taking account the measured QE of 0.25(2) for the APD at 906 nm, the in-fiber efficiencies for the source with and without stimulating pulse are extracted to be 0.03(2) and 0.016(2). We also estimated the extraction efficiency of the device by the using the APD counts and carefully calibrating all the loss in the setup, as shown in Fig.~S7. The estimated extraction efficiency of the device is $64.3/79.6 = 0.81$, which agrees very well with the theoretic value $\eta=F_p/(F_p+1)\times Q/Q_0=(6.6/7.6)\times(7548/8000)=0.82$. $F_p$ is the Purcell factor, Q is the quality factor of the cavity mode of the micropillar and $Q_0$ is the quality factor of the planar cavity. The brightness of our source could be dramatically boosted by just optimizing optics and QD quality. E.g., A transmission of $\sim$0.6 has been achieved in a similar setup~\cite{ding2016demand} by using high-transmission coating for NIR and near-unity QE of QDs has been demonstrated with a PIN gate~\cite{zhai2020low,schimpf2021entanglement}. An in-fiber efficiency as high as 0.81$\times$0.625=0.51 could be expected for non-blinking QD with optimized setup, as listed in the table of Fig.~S7.

\begin{figure}[!h]
	\includegraphics[width=\linewidth]{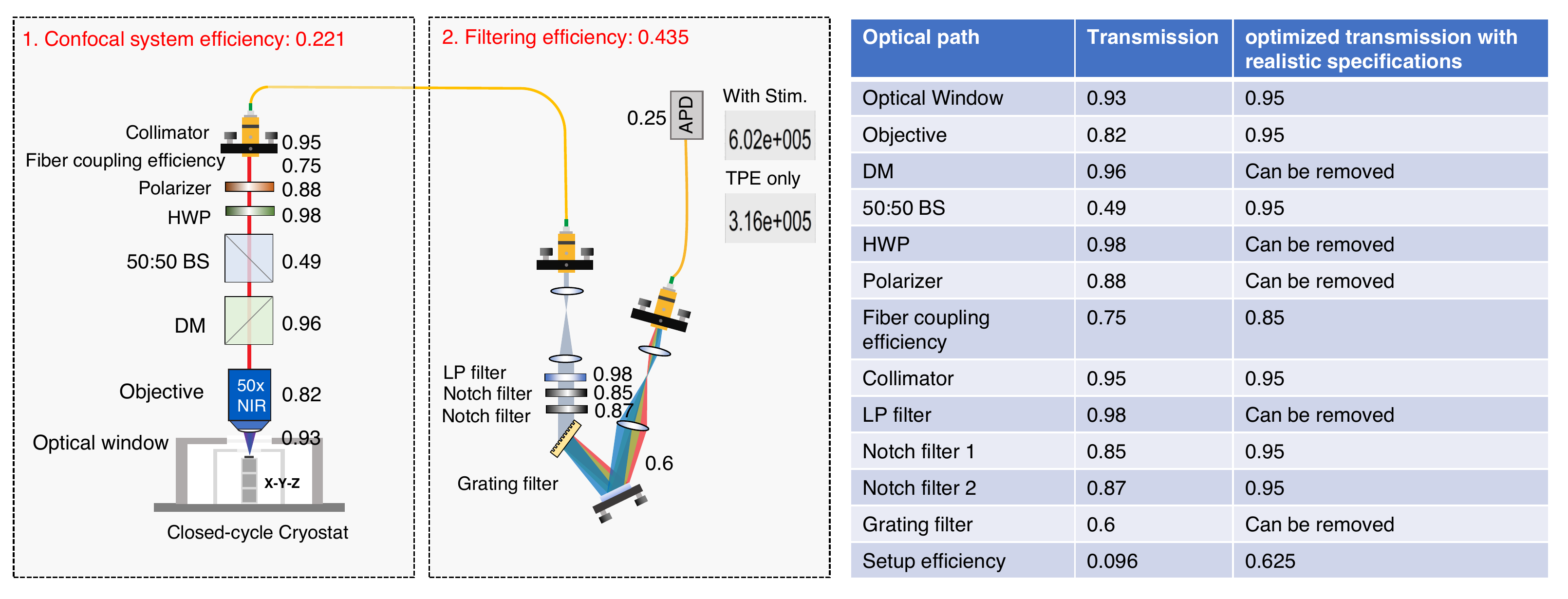}
	\caption{Calibration of the in-fiber source efficiency. Left: transmission of the all the optic elements in the setup. right table: the specifications of the optimized setup.}
	\label{SIfig:Fig8}
\end{figure}

\end{document}